\documentclass[epj]{svjour}
\usepackage{graphics}

\usepackage{mathrsfs}
\usepackage{bm}
\usepackage{bbm}
\usepackage{xr}
\usepackage{amsmath}
\usepackage{amssymb}
\usepackage{amsfonts}
\usepackage{graphicx}
\usepackage{lscape}

\usepackage[dvips]{color}

\providecommand*{\I}{\mathrm{i}}                           %% imaginary unit i
\providecommand*{\ket}[1]{|#1\rangle}                      %% ket vector
\providecommand*{\rbra}[1]{(#1|}                           %% bra with parenthesis
\providecommand*{\rket}[1]{|#1)}                           %% ket with parenthesis
\providecommand*{\lrbra}[1]{\rbra{\widetilde{#1}}}    %% left bra (with tilde) with parenthesis
\providecommand*{\rbracket}[2]{\rbra{#1}#2)}          %% bracket with parenthesis
\providecommand*{\lrbracket}[2]{\lrbra{#1}#2)}       %% bracket with left bra and right ket and parenthesis

\providecommand*{\klr}[1]{\left(#1\right)}					 %% variable sized parenthesis
\providecommand*{\mrmd}{d}									 %% differential ``d'' in mathrm
\newcommand{\mIm}{\mathrm{Im}}												%% imaginary part
\newcommand{\mRe}{\mathrm{Re}}												%% real part
\newcommand*{\mcal}[1]{\mathcal{#1}}                   %% shortcut for mathcal

\newcommand*{\umat}[1]{\underline{\mathscr{#1}}}      %% underline script symbol, used for matrices
\renewcommand{\vec}[1]{\bm{#1}}                       %% vectors are bold symbols here
\newcommand*{\uvec}[1]{\underline{\bm{#1}}}           %%  vectors with matrix components (vector operators)
\newcommand*{\vmc}[1]{\vec{\mathcal{#1}}}              %% vector in mathcal
\newcommand{\LambShift}{L}											%% Symbol for Lamb Shift
\newcommand{\FineStructure}{\Delta}										%% Symbol for fine structure
\newcommand{\HyperFineSplitting}{\mcal{A}}						%% Symbol for hyperfine splitting constant
\newcommand{\sez}{\vec{e}_0}													%% Symbol for spherical unit vector e_0
\newcommand{\sep}{\vec{e}_+}													%% Symbol for spherical unit vector e_+
\newcommand{\sem}{\vec{e}_-}													%% Symbol for spherical unit vector e_-

\newcommand{\mycell}[2]{\parbox{#1}{\vskip4pt #2 \vskip4pt}}	%% needed for tables in the appendix

\newcommand{\PV}{\mathrm{PV}}													%% subscript for PV quantities
\renewcommand{\thesection}{\arabic{section}}
\renewcommand{\thesubsection}{\arabic{section}.\arabic{subsection}}
\renewcommand{\thetable}{\arabic{table}}

\newcommand{\corresp}{\mathrel{\widehat{=}}}
\newcommand{\nn}{\nonumber}
\newcommand{\bx}{{\vec x}}

\newcommand{\mb}{\mathbf }
\begin{document}

\title{\Large Parity Violation in Hydrogen\\ and Longitudinal Atomic Beam Spin Echo I}

\author{Timo~Bergmann$^\mathrm{1,}\hspace{0.04em}$%
\thanks{Timo.Bergmann@web.DE}\fnmsep\thanks{present address: SAP AG, Walldorf}
\and
Maarten~DeKieviet$^\mathrm{2,}\hspace{0.04em}$%
\thanks{maarten@physi.Uni-Heidelberg.DE}
\and
Thomas~Gasenzer$^\mathrm{1,}\hspace{0.04em}$%
\thanks{T.Gasenzer@ThPhys.Uni-Heidelberg.DE}
\and
Otto~Nachtmann$^\mathrm{1,}\hspace{0.04em}$%
\thanks{O.Nachtmann@ThPhys.Uni-Heidelberg.DE}
\and
Martin-I.~Trappe$^\mathrm{1,}\hspace{0.04em}$%
\thanks{M.Trappe@ThPhys.Uni-Heidelberg.DE}
}

\institute{Institut f{\"u}r Theoretische Physik, Universit{\"a}t Heidelberg,\\ 
Philosophenweg 16, 69120 Heidelberg, Germany \and Physikalisches Institut, Universit{\"a}t Heidelberg,\\
Philosophenweg 12, 69120 Heidelberg, Germany}

\date{Received: \today}

\abstract{
We discuss the propagation of hydrogen atoms in static electric and magnetic fields in a longitudinal atomic beam spin echo (lABSE) apparatus. There the atoms acquire geometric (Berry) phases that exhibit a new manifestation of parity-(P-)violation in atomic physics. We provide analytical as well as numerical calculations of the behaviour of the metastable 2S states of hydrogen. The conditions for electromagnetic field configurations that allow for adiabatic evolution of the relevant atomic states are investigated. Our results provide the theoretical basis for the discussion of possible measurements of P-violating geometric phases in lABSE experiments.\\[2ex]
%\\[-7ex]
\hfill
{\small HD--THEP--08--19}
 % end of PACS codes
} %end of abstract
\titlerunning{Parity Violation in Hydrogen and longitudinal Atomic Beam Spin Echo I}
\authorrunning{T. Bergmann \textit{et al.}}
\maketitle

%========================================================================================
\section{Introduction}\label{s:Introduction}
%========================================================================================

In this paper we continue our theoretical studies towards the realisation of an experiment to measure P-violating effects using the methods available with an atomic beam interferometer as described in \cite{ABSE95}. Starting with the paper \cite{BoBo74} many theoretical and experimental studies of P-violation due to neutral current exchange were made for heavy atoms. For reviews see \cite{Khrip91,Bou97}. The most precise results so far are from experiments with Cs \cite{BeWi99,Vet95}. The experimental observation of P-violation in the lightest atoms, hydrogen (H) and deuterium (D), still is an open problem \cite{DuHo07}. For the early experimental efforts in this direction see \cite{LeWi75,HiHu77,AdFoTr81}. In \cite{BeGaNa07_I,BeGaNa07_II} we have studied parity violating geometric phases. These are interesting quantum mechanical phenomena by themselves and we think it is worthwhile to try measuring them. In the present paper we shall develop the necessary theoretical tools for a description of the behaviour of an atom in the interferometer described in \cite{ABSE95}. The aim is to calculate the relevant phases which atoms pick up when travelling through the interferometer and to construct observables sensitive to P-violating geometric phases. We shall focus on the hydrogen states with principal quantum number $n=2$, in particular on the metastable 2S states. Our paper is organised as follows. In Section \ref{s:HinInterferometer} we set up the basic equations governing the motion of the atom in the $n=2$ states as it travels through the interferometer. In Section \ref{s:SolutionOneComp} we study the equation for a single substate. Section \ref{s:InterferenceEffects} deals with two metastable 2S states and with their interference effects. In Section \ref{s:Conclusions} we draw our conclusions. A concrete proposal for a measurement of P-violating geometric phases will be given in a follow up paper. Our notation is explained in Appendix \ref{s:AppendixA} which collects also many useful formulae. Furthermore, we discuss in Appendix \ref{s:AppendixA} the conditions for adiabatic evolution of the atomic states for the case considered. Appendix \ref{s:AppendixB} contains the details of our calculations concerning the motion of a single 2S state in the interferometer. In Appendix \ref{s:AppendixC} we discuss in detail the interference effects of two states.
Throughout our work we use natural units $\hbar=c=1$ if not indicated otherwise.

%========================================================================================
\section{A hydrogen atom in the interferometer}\label{s:HinInterferometer}
%========================================================================================

\begin{figure}[htb]
\centering
\includegraphics[scale=0.485]{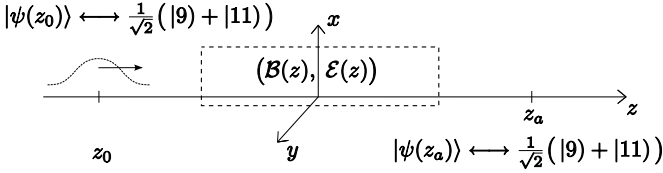}
\caption{Schematic view of the longitudinal spin echo interferometer. The coordinate system used is indicated. The coordinate along the beam direction is denoted by $z$. The atom is prepared in a wave packet around $z_0$ and analysed around $z_a$. In the example discussed in Section \ref{s:InterferenceEffects} we start with a superposition $\ket{\psi(z_0)}$ of two states, labeled with $\rket{9}$ and $\rket{11}$. After passing the magnetic and electric fields $\vec{\mathcal B}(z)$ and $\vec{\mathcal E}(z)$ the wavefunction is projected onto an analysing state $\ket{\psi(z_a)}$, for example again onto a superposition of the states $\rket{9}$ and $\rket{11}$.}
\label{SchematiclABSE}
\end{figure}

We consider in this section the case of a hydrogen atom being in a superposition of $n=2$ internal states and travelling through an interferometer of the type presented in \cite{ABSE95}. The latter is, for a theorist, a device producing static electric and magnetic fields acting on the atom. Schematically this is shown in Figure \ref{SchematiclABSE}.
The magnetic field strength $\vec{\mathcal B}$ and the electric field strength $\vec{\mathcal E}$ are supposed to depend only on the $z$ coordinate and to be nonzero only between $z_0$ and $z_a$. Here $z_0$ marks the place at the beginning of the interferometer around which, in a field-free region, the wave packet of the atom is prepared. Similarly, $z_a$ at the end of the interferometer marks the place where, again in a field-free region, the internal state of the atom is analysed.

The hydrogen atom in a superposition of states with principal quantum number $n=2$ is described by a 16-component spinor
\begin{align}\label{2.1}
|\psi(\mathbf x,t)\rangle =\sum^{16}_{\alpha=1}f_\alpha(\mathbf x,t)\rket{\alpha;0}\ .
\end{align}
Here $\rket{\alpha;0}$, $\alpha=1,\dots,16$, are the basis states for $\vec{\mathcal B} =0$ and $\vec{\mathcal E}=0$, that is the 2S$_{1/2}$, 2P$_{1/2}$ and 2P$_{3/2}$ states with the numbering as given in Table \ref{t:state.labels} of Appendix \ref{s:AppendixA}. The Schr\"odinger equation describing the undecayed $n=2$ states reads
\begin{align}\label{2.2}
\I\frac{\partial}{\partial t}|\psi(\mathbf x, t)\rangle=
\left[-\frac{1}{2m}\Delta +{\umat{M}}(z)+E_0\right]|\psi(\mathbf x,t)\rangle\ .
\end{align}
Here $m$ is the atom's mass and $\umat{M} (z)$ is the non-hermitian energy or mass matrix as described in Section \ref{II.s:Hydrogen} of \cite{BeGaNa07_II}. There the center of the real parts of the energy levels of the free 2P$_{1/2}$ states was chosen as the (arbitrary) zero for the energy scale. In the following it is convenient to choose this zero point of the energy differently. Therefore, we introduce in (\ref{2.2}) a real constant $E_0$ which we can choose freely. The mass matrix $\umat{M}(z)$ is given by
\begin{align}\label{2.4a}
\umat{M}(z)&=\umat{\tilde M}_0(\delta_1,\delta_2)-{\uvec D}\cdot {\vec{\mathcal E}}(z)-\uvec{\mu}\cdot {\vec{\mathcal B}}(z)\ ,\\
\umat{\tilde M}_0(\delta_1,\delta_2)&=\umat{M}_0+\delta_1\umat{M}^{(1)}_{\mathrm PV}
+\delta_2\umat{M}^{(2)}_{\mathrm PV}\label{2.4b}\ .
\end{align}
Here we use the same notation as in (\ref{e2:MM.full})--(\ref{e2:MM.free}) of \cite{BeGaNa07_II}. The mass matrix for zero external fields and without P-violating contributions is denoted by $\umat{M}_0$. The P-violating parts of the mass matrix are $\delta_i\umat{M}^{(i)}_{\mathrm PV}~,~i=1,2$. The two P-violation parameters $\delta_i$ quantifying the nuclear spin dependent and independent parts, respectively, are defined in Appendix \ref{s:AppendixA}. The matrices $\uvec D$ and $\uvec \mu$ are the electric and magnetic dipole-moment operators for the $n=2$ states. The explicit form of all these matrices is given in Appendix \ref{s:AppendixD}.

Let the momentum of the atom be
\begin{align}\label{2.5}
\vec q&=\left(\begin{array}{l}
\vec q_T\\q_z
\end{array}\right)\ ,\nn\\
\vec q_T&=\left(\begin{array}{l}
q_x\\q_y
\end{array}\right) \ .
\end{align}
The problem is to discuss the solution of (\ref{2.2}) for the case that the atom travels essentially in $z$ direction where for all $z$ the inequality
\begin{align}\label{2.5a}
\frac{q^2_z}{2m}\gg \lVert\umat{M}(z)\rVert
\end{align}
holds. Here $\lVert\umat{M}(z)\rVert$ is a suitable norm of $\umat{M}(z)$
\begin{align}\label{2.6}
\lVert\umat{M}(z)\rVert=\left[\mathrm{tr}\left(\umat{M}^\dagger (z)\umat{M}(z)\right)\right]^{1/2}\ .
\end{align}
Similarly, the kinetic energy of transverse motion is always supposed to be much smaller than that of longitudinal motion
\begin{align}\label{2.7}
\frac{q^2_z}{2m}\gg\frac{{\mb q}^2_T}{2m}\ .
\end{align}
Furthermore we suppose $\umat{M}(z)$ to vary only smoothly with $z$, such that the conditions for adiabatic evolution of the relevant states are fulfilled. This is specified for a concrete example below, and we discuss the adiabaticity criterion in detail in Appendix \ref{s:AppendixA}. For $z$ around $z_0$ and around $z_a$ (see Figure \ref{SchematiclABSE}) we suppose absence of external fields, thus 
\begin{align}\label{8aa}
\vec{\mathcal E}(z)=0\,,\; \vec{\mathcal B}(z)=0
\end{align}
for
\begin{align}\label{8a}
|z-z_0|<\delta l\; \mbox{ and } \;|z-z_a|<\delta l\ .
\end{align}
Here $\delta l$ is supposed to be very small compared to $z_a-z_0$,
\begin{align}
0<\delta l \ll z_a-z_0\ .
\end{align}
We have then
\begin{align}\label{2.8}
\umat{M}(z)=\umat{M}(z_0)=\umat{M}(z_a)=\umat{\tilde M}_0(\delta_1,\delta_2)\
\end{align}
for $|z-z_0|<\delta l$ and $|z-z_a|<\delta l$.

The next step is to introduce the right and left eigenvectors of the non-hermitian matrix $\umat{M} (z)$. We assume here that $\umat{M}(z)$ can be diagonalised for all $z$. For each $z$ we have then 16 linearly independent right and, analogously, left eigenvectors
\begin{align}\label{2.9}
\umat{M}(z)\rket{\alpha(z)}&=E_\alpha(z)\rket{\alpha(z)}\ ,\nn\\
\lrbra{\alpha(z)}\umat{M}(z)&=\lrbra{\alpha(z)}E_\alpha(z)\ ,\nn\\
\alpha&=1,\dots,16\ .
\end{align}
The eigenvalues $E_\alpha(z)$ are in general complex. We normalise the eigenvectors such that
\begin{align}\label{2.10}
\lrbracket{\alpha(z)}{\beta(z)}=\delta_{\alpha\beta}
\end{align}
and
\begin{align}\label{2.11}
&\rbracket{\alpha(z)}{\alpha(z)}=1\ ,\\
&\textup{(no summation over $\alpha$)}\ .\nn
\end{align}
We can now expand $|\psi(\mb x, t)\rangle$ in terms of $\rket{\alpha(z)}$
\begin{align}\label{2.12}
|\psi(\mb x,t)\rangle=\sum^{16}_{\alpha=1}\psi_\alpha(\mb x,t)\rket{\alpha(z)}\ .
\end{align}
From (\ref{2.2}) we get then
\begin{align}\label{2.13}
\sum^{16}_{\beta=1}&\left\{\left[-\frac{1}{2m}\Delta\psi_\beta(\vec x,t)
+\left(E_\beta(z)+E_0\right)\psi_\beta(\vec x,t)\right.\right.\nn\\
&\quad\left.-\I\frac{\partial\psi_\beta}{\partial t}(\vec x,t)\right]{|\beta(z))}\nn\\
&\quad-\frac1m\left(\frac{\partial\psi_\beta}{\partial z}(\vec x,t)\right)\frac{\partial}{\partial z}
{|\beta(z))}\nn\\
&\quad\left.-\frac{1}{2m}\psi_\beta(\vec x,t)\frac{\partial^2}{\partial z^2}{|\beta(z))}\right\}=0\ .
\end{align}
Multiplying with $\lrbra{\alpha(z)}$ from the left gives $16$ coupled equations
\begin{align}\label{2.14}
&-\frac{1}{2m}\Delta\psi_\alpha(\vec x,t)+\left(E_\alpha(z)+E_0\right)\psi_\alpha(\vec x,t)
-\I\frac{\partial\psi_\alpha}{\partial t}(\vec x,t)\nn\\
&+\sum^{16}_{\beta=1}\left\{-\lrbra{\alpha(z)}\frac{\partial}{\partial z}
|\beta(z))\frac1m\frac{\partial\psi_\beta}{\partial z}(\vec x,t)
\right.\nn\\
&\hspace{3.5em}\left.-\lrbra{\alpha(z)}\frac{\partial^2}{\partial z^2}|\beta(z))\frac{1}{2m}
\psi_\beta(\vec x,t)\right\}=0\ ,\nn\\
&\hspace{13.0em}(\alpha=1,\dots,16)\ .
\end{align}

In the following we shall consider (\ref{2.14}) in the adiabatic limit. That is, we want to choose conditions where the evolution of the metastable 2S states $(\alpha=9,\dots,12)$ decouples from that of the 2P states, see \cite{BeGaNa07_I,BeGaNa07_II}. This gives restrictions on the allowed rate of change of $\vec{\mathcal E}(z)$ and $\vec{\mathcal B}(z)$ with $z$. In order to get physical insight about these restrictions we shall now analyse them for a concrete case. Of course, the methods and formulas given below are easily adapted to other cases. %Therefore, we give now some order of magnitude estimates indicating when these conditions are indeed fulfilled.

%The following should be considered as an \textit{example} only. \textcolor{red}{Our methods are of course applicable to many other configurations, for which the adiabaticity conditions are met. In any case, the formulae given in the following sections give explicitly the probability of the atom to arrive at the end of the interferometer. This can then be used to decide if for a concrete experimental situation enough atoms are available for an analysis.}

\begin{figure}[htb]
\centering
\includegraphics[scale=1.06]{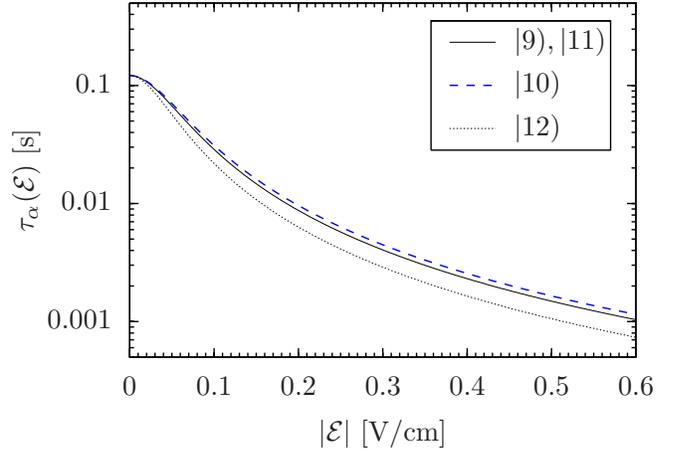}
\caption{Lifetimes of the mixed 2S$_{1/2}$ states of hydrogen in the external electric field $\vec{\mathcal E}=\mathcal E\vec{e}_3$ as function of $|\mathcal E|$. The numbering of the states ($\alpha=9,\dots,12)$ is explained in Table \ref{t:state.labels}, see Appendix \ref{s:AppendixA}.}
\label{Lifetimes}
\end{figure}

In the following example, the interferometer has a length of
\begin{align}\label{2.15}
l=z_a-z_0\approx 5\,\mathrm m\ .
\end{align}
The typical longitudinal velocity $v_z$, wave number $q_z$ and de Broglie wavelength $\lambda$ of  the H-atom in the interferometer at room temperature, $300\,$K, are
\begin{align}\label{2.16}
v_z&=\frac{q_z}{m}\approx 3500 \,\mathrm{m/s}\ ,\nn\\
q_z&\approx5.6\times10^{10}\,\mathrm m^{-1}\ ,\nn\\
\lambda&=\frac{2\pi}{q_z}\approx 1.1\times 10^{-10}\,\mathrm m\ .
\end{align}
This gives a kinetic energy of longitudinal motion
\begin{align}\label{2.17}
\frac{q^2_z}{2m}\corresp 6.5\times 10^{-2}\,\mathrm{eV}\
\end{align}
and a time of flight of the atom through the interferometer of
\begin{align}\label{2.17b}
\frac{l}{v_z}=1.43\,\mathrm{ms}\ .
\end{align}
The lifetimes of the free 2S and 2P states are \cite{LaShSo05,Sap04}
\begin{align}\label{2.18}
\tau_S&=\Gamma^{-1}_S=0.1216\,\mathrm s\ ,\nn\\
\tau_P&=\Gamma^{-1}_P=1.596\times 10^{-9}\,\mathrm s\ .
\end{align}
In an electric field the 2S states obtain a 2P admixture and their lifetimes decrease depending on the strength of $\vec{\mathcal E}$, see Figure \ref{Lifetimes}.
To have an appreciable flux of metastable 2S states at the end of the interferometer one should in view of (\ref{2.17b}) require, at least on average, 
\begin{align}\label{2.19}
\tau_S({\cal E})\gtrsim 1\,\mathrm{ms}\ .
\end{align}
From Figure \ref{Lifetimes} we see that this limits the allowed electric fields on average to 
\begin{align}\label{2.20}
\left|\vec{\mathcal E}(z)\right|_{av}\leq 0.60 \,\mathrm{V/cm}\ .
\end{align}
Of course, locally the fields can be stronger, but in practice a reasonable limit seems to be
\begin{align}\label{2.21}
\left|\vec{\mathcal E}(z)\right|_{max}\leq 10 \,\mathrm{V/cm}\ .
\end{align}

Estimates for the allowed rate of change of $\vec{\mathcal E}(z)$ and $\vec{\mathcal B}(z)$ compatible with adiabatic evolution of the atomic states are discussed in Appendix \ref{s:AppendixA} based on the results of \cite{BeGaNa07_II}. We find that the evolution of the 2S states in the interferometer will decouple from that of the 2P states if the typical length scale $\Delta z$ for a significant variation of the electric field $\vec{\mathcal E}(z)$ is $\Delta z\gtrsim 50\,\mathrm{\mu m}$. For the magnetic field $\vec{\mathcal B}(z)$ the length scale of significant variation should satisfy $\Delta z \gtrsim 10\,\mathrm{\mu m}$, see (\ref{A.110})--(\ref{A.117}) of Appendix \ref{s:AppendixA}.

\begin{figure}[htb]
\centering
\includegraphics[scale=0.95]{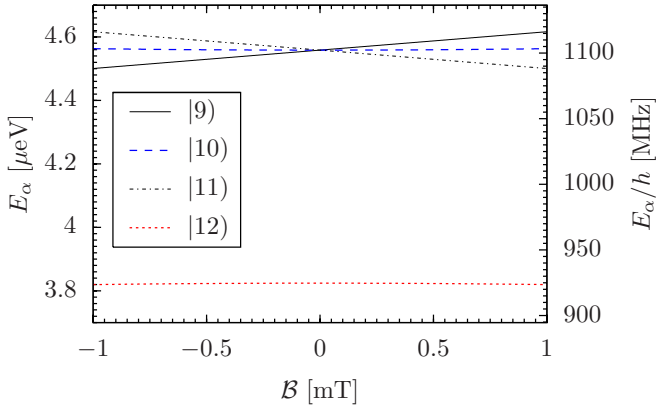}
\caption{Breit-Rabi diagram for the 2S states of hydrogen. The magnetic field is $\vec{\mathcal B}={\mathcal B}{\vec e}_3$.}
\label{BreitRabi}
\end{figure}

The occurrence of transitions amongst the individual 2S hyperfine states $\alpha=9,\dots,12$ in the interferometer depends very much on the precise setup of magnetic and electric fields since for $\vec{\mathcal B}(z)=0$ and $\vec{\mathcal E}(z)=0$ the states $\alpha=9,10$ and $11$, that is, the states 2S$_{1/2}$ with $F=1$ and $F_3=\pm 1,0$, are degenerate. The Breit-Rabi diagram for the states $\alpha=9,\dots,12$ is shown in Figure \ref{BreitRabi}. For concrete cases one must check the adiabaticity conditions explicitly. Typically the requirement is that the energy differences $\Delta E$ in the Breit-Rabi diagram for the $\vec{\mathcal B}$-fields considered correspond to frequencies larger than the frequencies $1/\Delta t$ of variation of the $\vec{\mathcal E}$- and $\vec{\mathcal B}$-fields. Setting $\Delta z=v_z\Delta t$ for the corresponding variation in $z$ we get
\begin{align}\label{2.22}
\Delta E\gtrsim\frac{hv_z}{\Delta z}\quad\Rightarrow\quad \Delta z\gtrsim\frac{v_z}{\Delta E/h}\ .
\end{align}
In this way we get, for example, for $\vec{\mathcal B}={\mathcal B}{\vec e}_3$ with 
$\mathcal B=0.5\,\mathrm{mT}$ for the $F=1$ states from Figure \ref{BreitRabi} $\Delta E\approx 0.05\,\mathrm{\mu eV}$, $\Delta E/h\approx 12\, \mathrm{MHz}$, $\Delta z\gtrsim 300\,\mathrm{\mu m}$. Of course, for other regions of the Breit-Rabi diagram one can make corresponding estimates.

Thus, we see that for the above example the conditions for adiabatic evolution of the 2S states lead to experimentally realisable restrictions for the electric and magnetic fields. With this we end our discussion of the example here, but we shall return to it in later sections.

Having discussed exemplarily the constraints on the electromagnetic field configuration for the adiabatic approximation to hold, we continue with the general investigation of (\ref{2.14}). In the following we suppose that the conditions for adiabatic evolution of the meta\-stable 2S states are fulfilled. Then, the evolution of these 2S states is governed by four decoupled equations obtained from (\ref{2.14}) for $\alpha=9,\dots,12$ and keeping there in the sum over $\beta$ only the terms with $\beta=\alpha$. In this way we get as the basic system of equations for the following discussions
\begin{align}\label{2.31}
&-\frac{1}{2m}\Delta\psi_\alpha(\vec x,t)
-\frac{1}{m}\lrbra{\alpha(z)}\frac{\partial}{\partial z}\rket{\alpha(z)}
\frac{\partial\psi_\alpha}{\partial z}(\vec x,t)\nn\\
&\quad+\left[E_\alpha(z)+E_0-\frac{1}{2m}\lrbra{\alpha(z)}
\frac{\partial^2}{\partial z^2}\rket{\alpha(z)}\right]\psi_\alpha(\vec x,t)\nn\\
&\quad-\I\frac{\partial}{\partial t}\psi_\alpha(\vec x,t)=0\ ,\nn\\
&\hspace{12.0em}(\alpha=9,\dots,12)\ .
\end{align}
To write (\ref{2.31}) in a convenient way we introduce the diagonal geometric phases (see (\ref{II.e3:BP.D}) of \cite{BeGaNa07_II})
\begin{align}\label{2.32}
\gamma_{\alpha\alpha}(z)=\I\int^z_{z_0}\mrmd z'\,
\lrbra{\alpha(z')}\frac{\partial}{\partial z'}\rket{\alpha(z')}\ ,\; (\alpha=9,\dots,12)\ .
\end{align}
Note that for metastable states $\gamma_{\alpha\alpha}(z)$ is, in general, complex. We split off from the wave function a factor corresponding to this geometric phase and define
\begin{align}\label{2.33}
\hat\psi_\alpha(\vec x,t)=\exp
[-\I\gamma_{\alpha\alpha}(z)]\psi_\alpha(\vec x,t)\ .
\end{align}
For $\hat\psi$ we find from (\ref{2.31}) the equation
\begin{align}\label{2.34}
\Delta\hat\psi_\alpha(\vec x,t)&+2\I m\frac{\partial\hat\psi_\alpha}{\partial t}(\vec x,t)-2m\mathcal V_\alpha(z)\hat\psi_\alpha(\vec x,t)=0\
\end{align}
where we define
\begin{align}\label{2.35}
\mathcal V_\alpha(z)&=E_\alpha(z)+E_0-\frac{1}{2m}\Bigg[\left(\frac{\partial\gamma_{\alpha\alpha}(z)}{\partial z}\right)^2
+\I\frac{\partial^2\gamma_{\alpha\alpha}(z)}{\partial z^2}\nn\\
&\hspace{10em}+\lrbra{\alpha(z)}\frac{\partial^2}{\partial z^2}\rket{\alpha(z)}\Bigg]\nn\\
&=E_\alpha(z)+E_0\nn\\
&\quad -\frac{1}{2m}\left[\left(\frac{\partial\gamma_{\alpha\alpha}(z)}{\partial z}\right)^2-\frac{\partial\lrbra{\alpha(z)}}{\partial z}\frac{\partial\rket{\alpha(z)}}{\partial z}\right]\ .
\end{align}
In the following we set
\begin{align}
\mathcal V_\alpha(z)&=V_\alpha(z)-\frac{\I}{2}\Gamma_\alpha(z)\ ,\nn\\
V_\alpha(z)&=\mRe\mathcal V_\alpha(z)\ ,\nn\\
\Gamma_\alpha(z)&=-2\mIm \mathcal V_\alpha(z)\ . \label{2.36}
\end{align}
The effective potential $\mathcal V_\alpha(z)$ is invariant under local phase transformations of the eigenstates $\rket{\alpha(z)}$ according to
\begin{align}\label{2.35b}
\rket{\alpha(z)}\longrightarrow\rket{\alpha(z)}'&=e^{\I\eta_\alpha(z)}\rket{\alpha(z)}\ ,\nn\\
\lrbra{\alpha(z)}\longrightarrow\lrbra{\alpha(z)}'&=e^{-\I\eta_\alpha(z)}\lrbra{\alpha(z)}
\end{align}
where $\eta_\alpha(z)$ has to be real in order to respect (\ref{2.11}), see Appendix \ref{s:AppendixB}.

The solutions of (\ref{2.34}) will be discussed in the next sections using methods from wavelet theory, see for instance \cite{FT98}.

%---------------------------------------------------------------------------------------------------
\section{Solution for one component}\label{s:SolutionOneComp}
%---------------------------------------------------------------------------------------------------

We now study the solution of the effective equation (\ref{2.34}) for the amplitudes of the metastable states for one component $\alpha$. Since we keep $\alpha$ fixed in this section we omit this index from our quantities in the following. Thus, we drop the index $\alpha$ in (\ref{2.33})
\begin{align}\label{3.0}
\hat\psi(\vec x,t)=\exp[-\I\gamma(z)]\psi(\vec x,t)\ ,
\end{align}
and the equation to be studied for $\hat\psi$ reads
\begin{align}\label{3.1}
\bigg\{
\frac{\partial^2}{\partial z^2}+2\I m\frac{\partial}{\partial t}
&-2mV(z)\nn\\
&+\I m\Gamma(z)
+\Delta_T\bigg\}\hat\psi(\vec x,t)=0
\end{align}
with
\begin{align}\label{3.2}
\vec x&=\left(\begin{array}{c}\vec x_T\\z\end{array}\right)=\left(\begin{array}{c}x\\ y\\ z\end{array}\right)\ ,\nn\\
\Delta_T&=\frac{\partial^2}{\partial x^2}+\frac{\partial^2}{\partial y^2}\ .
\end{align}
Let $\bar k$, of order $q_z$, be a typical scale for the momentum component in $z$ direction. We change variables and set
\begin{align}\label{3.3}
\tau&=\frac{\bar k}{m}t\ ,\nn\\
\zeta&=\int^z_{z_0}\mathrm dz'\frac{\bar k}{\sqrt{\phantom{^|}\bar k^2-2mV(z')}}\ .
\end{align}
From (\ref{3.3}) we get $\zeta$ as function of $z$. For this and for the inverse function we write
\begin{align}\label{3.3a}
\zeta&=\frak{Z}(z)\ ,\nn\\
z&=\mathcal Z(\zeta)\ .
\end{align}
Furthermore, we split off from the wave function $\hat\psi(\vec x,t)$ a convenient factor $\exp[\I\phi(z,t)]$ and denote by $\mathcal A$ the remaining amplitude which is the envelope function for the wave packet:
\begin{align}\label{3.4}
\hat\psi(\vec x,t)=e^{\I\phi(z,t)}\mathcal A(\vec x_T,\zeta,\tau)\ .
\end{align}
In the following we could choose $\phi$ to be the WKB phase factor
\begin{align}\label{3.5}
\phi_{\mathrm{WKB}}(z,t)=-\frac{\bar k^2}{2m}t+\int^z_{z_0}\mathrm dz'\, k(z')\ .
\end{align}
Here, the local momentum $k(z)$ is defined as
\begin{align}\label{3.6}
k(z)=\sqrt{\bar k^2-2mV(z)}\ .
\end{align}
Then $\exp[\I\phi_{\mathrm{WKB}}(z,t)]$ is a pure phase factor since $\phi_{\mathrm{WKB}}$ is real. However, taking into account a simple kinematic effect in the wave function when $k(z)$ changes as well as the decay of the metastable atom, we rather choose a complex $\phi$
\begin{align}\label{3.7}
\phi(z,t)=\left\{ -\frac{\bar k^2}{2m}t\right.+\int_{z_0}^z&\mathrm dz'\,k(z')+\frac{\I}{2}\ln\frac{k(z)}{k(u)}\nn\\
&\quad\left.\left.+\frac{\I}{2}\int_u^z\mathrm dz'\frac{m\Gamma(z')}{k(z')}\right\}\right|_{u=u(z,t)}\nn\\
&\
\end{align}
where
\begin{align}\label{3.8}
u(z,t)&=\mathcal Z(\zeta-\tau)\nn\\
&=\mathcal Z\left({\frak{Z}}(z)-\frac{\bar k}{m}t\right)\ .
\end{align}

It is now a simple exercise to transform (\ref{3.1}) to the new variables $\vec x_T,\zeta,\tau$ and to derive the equation for the envelope function ${\mathcal A}(\vec x_T,\zeta,\tau)$ of (\ref{3.4}). For the details see Appendix \ref{s:AppendixB}. The result can be written in the form
\begin{align}\label{3.9}
\left(\frac{\partial}{\partial\tau}+\frac{\partial}{\partial\zeta}\right)
\mathcal A(\vec x_T,\zeta,\tau)=(L\mathcal A)(\vec x_T,\zeta,\tau)
\end{align}
where $L$ is an operator given in detail in (\ref{B.16})--(\ref{B.19}). The initial condition for ${\mathcal A}$ is 
\begin{align}\label{3.9a}
\left.\mathcal A(\vec x_T,\zeta,\tau)\right|_{\tau=0}=\varphi(\vec x_T,\zeta)
\end{align}
with some given function $\varphi$ determined by the initial experimental conditions. With this the solution of (\ref{3.9}) is easily obtained by transforming (\ref{3.9}) into an integral equation. Here we use a Green's function method, see Appendix \ref{s:AppendixB}. The final result is an expansion of ${\mathcal A}$, see (\ref{B.32}),
\begin{align}\label{3.10}
{\mathcal A}(\vec x_T,\zeta,\tau)=\sum^\infty_{n=0}{\mathcal A}^{(n)}
(\vec x_T,\zeta,\tau)
\end{align}
where successive terms are suppressed by higher and higher powers of $1/\bar k$
\begin{align}\label{3.11}
{\mathcal A}^{(n)}(\vec x_T,\zeta,\tau)=\mathcal O(1/\bar k^n)\ .
\end{align}
For large $\bar k$ the zero order term should be a good approximation. It reads, see (\ref{B.33}),
\begin{align}\label{3.12}
{\mathcal A}^{(0)}(\vec x_T,\zeta,\tau)=\varphi(\vec x_T,\zeta-\tau)\ .
\end{align}
The higher order terms in (\ref{3.10}) are discussed in Appendix \ref{s:AppendixB}.
The zero order term (\ref{3.12}) describes the motion of the wave packet without longitudinal and transverse dispersion. The envelope function is just progressing with constant speed in the $\zeta$ coordinate, where $\zeta={\frak{Z}}(z)$, see (\ref{3.3}) and (\ref{3.3a}).

To discuss the physics of the solution (\ref{3.12}) let us suppose that for $t=0$ the envelope function $\varphi(\vec x_T,{\frak{Z}}(z))$ is sharply peaked at $\vec x_T=0$ and $z=z_0$ corresponding to $\zeta=0$. Then, according to (\ref{3.12}), the envelope function for times $t>0$ stays equally peaked at $\vec x_T=0$ but the peak in $z$ moves according to $z(t)$ where
\begin{align}\label{3.13}
\frak{Z}(z(t))-\frac{\bar k}{m}t=\frak{Z}(z_0)=0\ .
\end{align}
Inserting here $\frak{Z}(\cdot)$ from (\ref{3.3}), (\ref{3.3a}) we get
\begin{align}\label{3.14}
\int^{z(t)}_{z_0}\mathrm dz'
\frac{m}{\sqrt{\phantom{^|}\bar k^2-2mV(z')}}=t
\end{align}
which gives
\begin{align}\label{3.15}
\frac12 m\dot{z}^2(t)+V(z(t))=\frac{\bar k^2}{2m}\ .
\end{align}
The motion of the peak in $z$ corresponds to that of a classical particle in the real part of the potential where the total energy is $\bar k^2/2m$.

Now we can make contact with the roadmap description (``Fahrplanmodell''), introduced in \cite{DissAR} and in a manu\-script currently prepared for publication \cite{16a}. We consider wave packets starting at $t=0$ in the field free region at $z_0$, see Figure \ref{SchematiclABSE}. We define the arrival time $t(z)$ and the corresponding reduced ``time'' $\tau(z)$ by reversing the functional dependence in (\ref{3.14})
\begin{align}\label{3.16}
t(z)&=\int^z_{z_0}\mathrm dz'\frac{m}{\sqrt{\phantom{^|}\bar k^2-2mV(z')}}\ ,\nn\\
\tau(z)&=\frac{\bar k}{m}t(z)={\frak{Z}}(z)\ .
\end{align}
Note that the reduced ``time'' $\tau$ has the dimension of length. We compare these arrival times with those of a free wave packet where $V(z')\equiv 0$. From (\ref{2.35}) and (\ref{2.36}) we see that this corresponds to a free atom where $\gamma(z)\equiv 0$ and the energy level $E(z')$ satisfies
\begin{align}\label{3.16b}
\mathrm{Re}\,E(z)+E_0=0\ .
\end{align}
That is, the choice of $E_0$ selects the free state to which one compares, where the arrival time $t$ and reduced ``time'' $\tau$ are
\begin{align}\label{3.17}
t_{\mathrm{free}}(z)&=\frac{m}{\bar k}(z-z_0)\ ,\nn\\
\tau_{\mathrm{free}}(z)&=\frac{\bar k}{m}t_{\mathrm{free}}(z)=z-z_0\ .
\end{align}
The difference of reduced arrival times 
\begin{align}\label{3.18}
\Delta\tau(z)&=\tau(z)-\tau_{\mathrm{free}}(z)\nn\\
&={\frak{Z}}(z)-(z-z_0)
\end{align}
describes the delay $(\Delta\tau>0)$ or advance $(\Delta\tau<0)$ of the arrival of the wave packet peak at coordinate $z$ in the interferometer with respect to that of the free atom.

\begin{figure}[htb]
\centering
\includegraphics[scale=1.1]{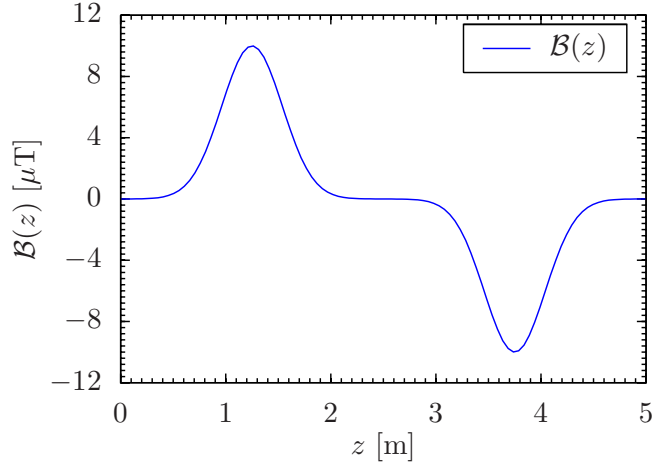}
\caption{Example of a magnetic field configuration. The transverse components are chosen as $\mathcal B_1=\mathcal B_2=0$ and the longitudinal component $\mathcal B(z)$ is given in (\ref{3.18a}).}
\label{B3Figure}
\end{figure}

\begin{figure}[htb]
\centering
\includegraphics[scale=1.1]{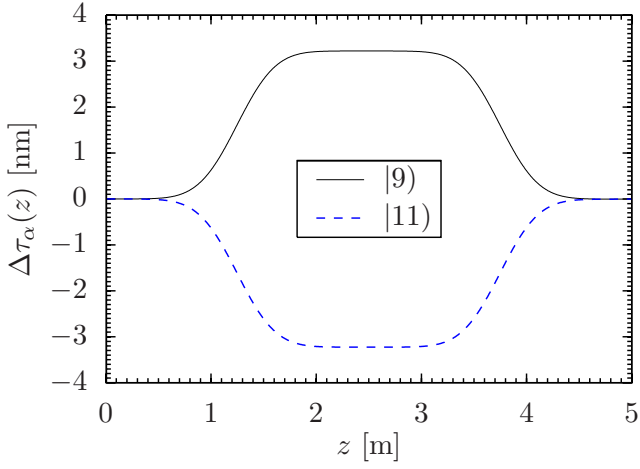}
\caption{Example of a roadmap plot for the states $\alpha=9$ and $\alpha = 11$. The corresponding potentials are given in (\ref{3.19}).}
\label{TimeTable}
\end{figure}

As a concrete example we consider again the interferometer with the parameters specified in (\ref{2.15}) ff. In Figure~\ref{TimeTable} the roadmap plot is shown for the states $\alpha=9$ and $\alpha=11$ in a purely longitudinal magnetic field
\begin{align}\label{3.18a}
\vec{\mathcal B}(z)&=\mathcal B(z)\vec e_3\ , \nn\\
\mathcal B(z)&=10\,\left(e^{-6(z-1.25)^2}-e^{-6(z-3.75)^2}\right)\,\mu\mathrm{T}\ ,
\end{align}
depicted in Figure \ref{B3Figure}, where $z$ is taken in meters.
We compare the evolution of these states with that of the free 2S, $F=1$, states ($\alpha=9$, $10$ and $11$). That is, we set
\begin{align}
E_0=-L-\frac{\mathcal A}{32}\ ,
\end{align}
see Table \ref{t:H2.M0} in Appendix \ref{s:AppendixD}. Here $\mathcal A$ denotes the ground state hyperfine splitting energy. The potentials for the states $\alpha=9$ and $\alpha=11$ are then proportional to the third component of the magnetic field in (\ref{3.18a}):
\begin{align}\label{3.19}
V_9(z)&=\frac{g\mu_B}{2}\mathcal B(z)\ ,\nn\\
V_{11}(z)&=-\frac{g\mu_B}{2}\mathcal B(z)\ ,
\end{align}
with the Bohr magneton $\mu_B$ and the electron's Land\'{e} factor $g$, see Table \ref{t:H2.Mu} in Appendix \ref{s:AppendixD}.
We see from Figure \ref{TimeTable} that in the magnetic field (\ref{3.18a}) the state $\alpha =9$ first gets a delay, $\alpha=11$ an advance compared to the free state. For $z$ around $3.75\,$m where the magnetic field is reversed the state $\alpha=9$ makes up for the delay, the state $\alpha=11$ for its advance. Finally at $z=5\,$m they meet again. This gives a nice visualisation of the motion of the wave packet in the interferometer. Clearly, the complete solution contains much more information, as discussed below.

Returning now to the general discussion and putting everything together we find from (\ref{3.0}), (\ref{3.4}), (\ref{3.7}) and (\ref{3.12}) the lowest order solution for the wave function as
\begin{align}\label{3.20}
\psi(\vec x,t;\bar k)&=\exp[\I\gamma(z)]
\hat\psi(\vec x,t)\nn\\
&=\exp[\I\phi(z,t)+\I\gamma(z)]
{\mathcal A}^{(0)}(\vec x_T,\zeta,\tau)\nn\\
&=\exp[\I\phi(z,t)+\I\gamma(z)]
\varphi(\vec x_T,\zeta-\tau)\ .
\end{align}
For purposes which will become clear in Section \ref{s:InterferenceEffects} we indicate here explicitly also the dependence of $\psi$ on $\bar k$. To write this in a physically transparent form we note that (see (\ref{3.6}))
\begin{align}\label{3.21}
k(z)-\bar k&=\frac{k^2(z)-\bar k^2}{k(z)+\bar k}=-\frac{2mV(z)}{k(z)+\bar k}\ ,\nn\\
\int^z_{z_0}\mathrm dz'k(z')&=\bar k(z-z_0)+\int^z_{z_0}\mathrm dz'(k(z')-\bar k)\nn\\
&=\bar k(z-z_0)-\int^z_{z_0}\mathrm dz'
\frac{2mV(z')}{k(z')+\bar k}\ .
\end{align}
We define now
\begin{align}\label{3.22}
&U(z,u;\bar k)=\exp\left[-\I\int^z_u\mathrm dz'\frac{2mV(z')}{k(z')+\bar k}\right.\nn\\
&\left.\qquad-\frac{1}{2}\ln\frac{k(z)}{k(u)}-\frac12\int^z_u\mathrm dz'\frac{m\Gamma(z')}{k(z')}+\I\gamma(z)-\I\gamma(u)\right]\ .
\end{align}
It is easy to see that $U$ corresponds to the evolution of the wave function of an atom at rest with an effective complex mass
\begin{align}\label{3.23}
\umat M_{\mathrm{eff}}(z)&=\frac{2k(z)}{k(z)+\bar k}V(z)+\frac{\I}{2k(z)}\frac{\partial V(z)}{\partial z}\nn\\
&\quad-\frac{\I}{2}\Gamma(z)-\frac{k(z)}{m}\frac{\partial\gamma(z)}{\partial z}\ .
\end{align}
We get
\begin{align}\label{3.24}
\frac{k(z)}{m}\frac{\partial U(z,u;\bar k)}{\partial z}&=-\I \umat M_{\mathrm{eff}}(z)U(z,u;\bar k)\ ,\nn\\
\left. U(z,u;\bar k)\right|_{z=u}&=1\ .
\end{align}
The solution (\ref{3.20}) can now be written as
\begin{align}\label{3.26}
\psi(\vec x,t;\bar k)&=
\exp\left[-\I\frac{\bar k^2}{2m}t+\I\bar k(z-z_0)\right]\nn\\
&\quad\times \left. U(z,u;\bar k)\right|_{u=\mathcal Z(\zeta-\tau)}\chi{(\vec x_\tau,\zeta-\tau)}
\end{align}
where we define
\begin{align}\label{3.27}
&\chi{(\vec x_T,\zeta-\tau)}=\nn\\
&\qquad\left.\exp\left[-\I\int^u_{z_0}\mathrm dz'\frac{2mV(z')}{k(z')+\bar k}+\I\gamma(u)\right]\right|_{u=\mathcal Z(\zeta-\tau)}\nn\\
&\qquad\quad\times \varphi(\vec x_T,\zeta-\tau)\ .
\end{align}
For $\tau=0$ we have $u=\mathcal Z(\zeta)=z$ and $U(z,u;\bar k)=1$. Thus,
\begin{align}\label{3.28}
\left.\psi(\vec x,t;\bar k)\right|_{t=0}=
\exp[\I\bar k(z-z_0)]\left.\chi(\vec x_T,\zeta)\right|_{\zeta={\frak{Z}}(z)}
\end{align}
contains a plane wave factor and $\chi(\vec x_T,\zeta)$, the envelope amplitude at $t=0$. The interpretation of the result (\ref{3.26}) is then as follows. The wave function $\psi(\vec x,t;\bar k)$ contains a plane wave factor and a factor $U(z,u)$ corresponding to the complex phase picked up by the atom travelling from $u=\mathcal Z(\zeta -\tau)$ to $z=\mathcal Z(\zeta)$. This is multiplied with the shifted original envelope amplitude $\chi(\vec x_T,\zeta-\tau)$. The factor $U(z,u;\bar k)$ corresponds to the evolution of the atom at rest with the complex mass $\umat M_{\mathrm{eff}}(z)$ (\ref{3.23}). From (\ref{3.22}) we see that $U$ contains, in general, complex dynamical and geometric phases. In this way we make the connection to \cite{BeGaNa07_I,BeGaNa07_II} where such phases were studied for atoms at rest.

A typical ansatz for the initial envelope amplitude $\chi$ in (\ref{3.26})--(\ref{3.28}) is a Gaussian
\begin{align}\label{3.29}
\chi(\vec x_T,\zeta)&={\cal N}(2\pi)^{-3/4}\sigma^{-1}_T\sigma^{-1/2}_L\nn\\
&\quad\times\exp \left(-\frac{1}{4\sigma^2_T}\vec x^2_T
-\frac{1}{4\sigma^2_L}\zeta^2\right)\ ,\nn\\
\sigma_T,\sigma_L&>0\ .
\end{align}
Here ${\cal N}>0$ is a normalisation factor, $\sigma_T$ the transverse and $\sigma_L$ the longitudinal width, respectively. Supposing that at $t=0$ we have exactly one atom in this wave we get ${\cal N}$ from 
\begin{align}\label{3.30}
\int \mathrm d^3x\left.|\psi(\vec x,t;\bar k)|^2\right|_{t=0}=\int \mathrm d^2x_T\int \mathrm dz\,|\chi(\vec x_T,{\frak{Z}}(z))|^2=1\ .
\end{align}

In Appendix \ref{s:AppendixB} we study the first correction term in the series (\ref{3.10}) for the function (\ref{3.29}). We find that for the interferometer parameters specified in (\ref{2.15}) ff. the zero order solution (\ref{3.26}) is valid with a $1\%$ accuracy in the region of interest provided
\begin{align}\label{3.31}
\sigma_L&>100\,\mathrm{\mu m}\ ,\nn\\
\sigma_T&>100\,\mathrm{\mu m}\ ,
\end{align}
see (\ref{B.55a})--(\ref{B.57}). Under typical experimental conditions \cite{ABSE95} the actual longitudinal widths of the wave packets used are of order $0.4\,$nm, that is, much smaller than those in (\ref{3.31}), while $\sigma_T$ can indeed be of order $100\,\mathrm{\mu m}$. Such wave packets are described as superpositions of the solutions (\ref{3.26}) which serve as wavelets in the analysis, see Section \ref{s:InterferenceEffects}.

%---------------------------------------------------------------------------------------------------
\section{Interference effects}\label{s:InterferenceEffects}
%---------------------------------------------------------------------------------------------------

In this section we shall study, as a concrete example, interference effects of two 2S states travelling through the interferometer specified in (\ref{2.15}) ff. Of course, the formulas given below are easily adapted to other experimental setups and more than two states.

Let us consider the case that we have at time $t=0$ two narrow wave packets around $z\approx z_0$ in the interferometer of (\ref{2.15}) ff. These wave packets are supposed to form a coherent superposition of two components $\rket{\alpha(z_0)}$ and $\rket{\beta(z_0)}$ of the 2S states where
\begin{align}\label{4.1}
\alpha,\beta\in I\subset\{9,10,11,12\}\ .
\end{align}
Here $I$ is defined as an index set with two elements. The two waves travel through the interferometer and are supposed to be analysed at $z=z_a$ as discussed below. For the initial state we make the following ansatz
\begin{align}\label{4.2}
\left.\ket{\Psi(\vec x,t)}\right|_{t=0}=
\Psi(\vec x,0)\sum_{\alpha\in I}c_\alpha\rket{\alpha(z_0)}
\end{align}
with
\begin{align}\label{4.3}
\sum_{\alpha\in I}|c_\alpha|^2=1
\end{align}
and
\begin{align}\label{4.3a}
\Psi(\vec x,0)
=\int\frac{\mathrm d\bar k}{2\pi}f(\bar k)
\exp[\I\bar k(z-z_0)]\chi(\vec x_T,z-z_0)\ .
\end{align}
Here and in the following all integrals run from $-\infty$ to $+\infty$ if not indicated otherwise. This initial wave function is supposed to be non zero only in the field free region around $z_0$. Taking into account (\ref{3.31}) we shall now suppose that the parameter $\delta l$ specifying the field free region in (\ref{8a}) has the value
\begin{align}\label{71a}
\delta l = 10\,\mathrm{mm}\ .
\end{align}
Note that for zero external fields the initial states $\rket{\alpha(z_0)}$ with $\alpha\in I$ are orthonormal in the ordinary sense since these states have different values $(F,F_3)$ which are ``good'' quantum numbers for $\vec{\mathcal E}=\vec{\mathcal B}=\vec 0$, see Table \ref{t:state.labels} of Appendix \ref{s:AppendixA}. In (\ref{4.3a}) $\chi(\vec x_T,z-z_0)$ is an envelope amplitude with widths in transverse and longitudinal directions
\begin{align}\label{71b}
\delta l = 10\,\mathrm{mm}\gg\sigma_T,\sigma_L\gtrsim 100\,\mathrm{\mu m}\ .
\end{align}
We suppose the normalisation condition
\begin{align}\label{4.4}
\int \mathrm d^2x_T\int \mathrm dz\,|\chi(\vec x_T,z-z_0)|^2=1
\end{align}
to hold, see (\ref{3.30}) with $\frak{Z}(z)=z-z_0$. The Fourier transform of $\chi$ is denoted by
\begin{align}\label{4.5}
\tilde{\chi}(\vec q_T,q_L)=\int \mathrm d^2x'_T\int \mathrm dz'
\exp(-\I\vec q_T&\vec x'_T-\I q_Lz')\nn\\
&\times\chi(\vec x'_T,z')\ .
\end{align}
We superpose functions $\chi$ in (\ref{4.3a}) using a function $f(\bar k)$ which is supposed to have a sharp peak at $\bar k=\bar k_m$ ($m$ for maximum) and a width $\Delta\bar k=\sigma^{-1}_k$ with
\begin{align}\label{4.5a}
0<\sigma_k\ll\sigma_L,\; \bar k_m\gg\sigma^{-1}_k\ .
\end{align} We require the state in (\ref{4.2}) to be normalised which implies
\begin{align}\label{4.6}
\int \mathrm d^3x|\Psi(\vec x,0)|^2 &=\int\frac{\mathrm d\bar k'\mathrm d\bar k}{(2\pi)^2}\,f^*(\bar k')f(\bar k)\nn\\
&\qquad\times\int \mathrm d^2x'_T\int \mathrm dz'
\exp [\I(\bar k-\bar k')z']\nn\\
&\qquad\times|\chi(\vec x'_T,z')|^2\nn\\
&=1\ .
\end{align}
We define the Fourier transform of $\Psi(\vec x,0)$ in (\ref{4.3a}) as 
\begin{align}\label{4.7}
&\tilde{\Psi}(\vec q_T,q_L)\nn\\
&\quad=\int \mathrm d^2x_T\int \mathrm dz\exp(-\I\vec q_T\vec x_T-\I q_L(z-z_0))
\Psi(\vec x,0)\nn\\
&\quad=\int\frac{\mathrm d\bar k}{2\pi}f(\bar k)\tilde\chi(\vec q_T,q_L-\bar k)\ .
\end{align}

Equations (\ref{4.1}) to (\ref{4.7}) define the general type of wave functions for which we shall calculate interference effects below. For later use, we define Gaussian functions as concrete examples for such wave functions:
\begin{align}
\chi(\vec x_T,z-z_0)&=(2\pi)^{-3/4}\sigma^{-1}_T\sigma^{-1/2}_L\nn\\ &\quad\times\exp\left[-\frac{1}{4\sigma^2_T}\vec x^2_T-\frac{1}{4\sigma^2_L}(z-z_0)^2\right],\label{4.8}\\
f(\bar k)&=(4\pi\sigma_k)^{1/2}(\sigma^2_k+\sigma^2_L)^{1/4}\nn\\
&\quad\times\exp\left[-\sigma^2_k(\bar k-\bar k_m)^2\right]\label{4.9}
\end{align}
with $10\,\mathrm{mm}\gg\sigma_T$, $\sigma_L\geq 100\,\mathrm{\mu m}$ and $\sigma_k\ll \sigma_L$. We get then
\begin{align}\label{4.10}
\tilde\chi(\vec q_T,q_L)&=(2\pi)^{3/4}2\sigma_T\sqrt{2\sigma_L}\nn\\
&\quad\times\exp(-\sigma^2_T\vec q^2_T-\sigma^2_Lq^2_L)\ ,\\
\tilde{\Psi}(\vec q_T,q_L)&=(2\pi)^{3/4}2\sigma_T\sqrt{2\sigma'_k}\nn\\
&\quad\times\exp[-\sigma^2_T\vec q^2_T-\sigma'^2_k(q_L-\bar k_m)^2]\label{4.11}
\end{align}
where 
\begin{align}\label{4.12}
\sigma'_k=\frac{\sigma_k\sigma_L}{\sqrt{\sigma^2_L+\sigma^2_k}}\ .
\end{align}
Note that for $\sigma_k\ll\sigma_L$ we have $\sigma'_k\approx\sigma_k$.

The evolution of the general state (\ref{4.2}) with time is obtained from the results of Section \ref{s:SolutionOneComp}. Here, of course, we have to reinstate the index $\alpha$ on all quantities. Thus we have potentials $\mathcal V_\alpha$, variables $\zeta_\alpha$, final solutions (\ref{3.26}) $\psi_\alpha(\vec x,t;\bar k)$, etc. We get in this way from the initial state (\ref{4.2}) the solution
\begin{align}\label{4.13}
|\Psi(\vec x,t)\rangle=\int\frac{\mathrm d\bar k}{2\pi}f(\bar k)
\sum_{\alpha\in I}c_\alpha\psi_\alpha(\vec x,t;\bar k)\rket{\alpha(z)}
\end{align}
where
\begin{align}\label{4.14}
\psi_\alpha(\vec x,t;\bar k)&=\exp\left[-\I\frac{\bar k^2}{2m}t+\I\bar k(z-z_0)\right]\nn\\
&\quad\times\left.U_\alpha(z,u;\bar k)\right|_{u=\mathcal Z_\alpha({\frak{Z}}_\alpha(z)-(\bar k/m)t)}\nn\\
&\quad\times\chi(\vec x_T,{\frak{Z}}_\alpha(z)-(\bar k/m)t)\ .
\end{align}
Now we come to the analysis of the state (\ref{4.13}) at $z=z_a$ where again we have a field free region for $z$ several $\sigma_L$ around $z_a$, see (\ref{8a}) and (\ref{71b}). We have, therefore,
\begin{align}\label{4.15}
\rket{\alpha(z)}=\rket{\alpha(z_a)}=\rket{\alpha(z_0)}\
\end{align}
for $|z-z_a|<\delta l =10 \,\mathrm{mm}$. We suppose that the state (\ref{4.13}) is projected onto
\begin{align}\label{4.16}
\rket{p}=\sum_{\alpha \in I}p_\alpha\rket{\alpha(z_0)}
\end{align}
where 
\begin{align}\label{4.17}
\sum_{\alpha\in I}|p_\alpha|^2=1\ .
\end{align}
This gives
\begin{align}\label{4.18}
\ket{\Psi_p(\vec x,t)}
=\rket{p}\lrbra{p}\,\Psi(\vec x,t)\rangle
=\Psi_p(\vec x,t)\rket{p}
\end{align}
with
\begin{align}\label{4.19}
\lrbra{p}=\sum_{\alpha\in I}\lrbra{\alpha(z_0)}p^*_\alpha\ ,
\end{align}
\begin{align}\label{4.20}
\Psi_p(\vec x,t)&=
\lrbra{p}\Psi(\vec x,t)\rangle\nn\\
&=\int\frac{\mathrm d\bar k}{2\pi}f(\bar k)
\sum_{\alpha\in I}p^*_\alpha c_\alpha\psi_\alpha(\vec x,t;\bar k)\ .
\end{align}

We suppose that the total integrated flux ${\cal F}_p$ in the above projection is measured at $z=z_a$. This is given by
\begin{align}\label{4.21}
{\cal F}_p=\int \mathrm dt\int \mathrm d^2x_T \,j_z(\vec x_T,z_a,t)
\end{align}
where the $z$-component of the probability current is 
\begin{align}\label{4.22}
j_z(\vec x_T,z,t)=
\frac{1}{2m\I}\Psi^*_p(\vec x,t)\frac{\partial}{\partial z}\Psi_p(\vec x,t)
+\mathrm{c.c.} \ .
\end{align}
In evaluating $j_z$ and ${\cal F}_p$ we take into account (\ref{4.5a}). We get then, as explained in detail in Appendix \ref{s:AppendixC}, the following results
\begin{align}\label{4.23}
j_z(\vec x_T,z_a,t)=
\frac{\bar k_m}{m}\Psi^*_p(\vec x_T,z_a,t)\Psi_p(\vec x_T,z_a,t)\ ,
\end{align}
\begin{align}\label{4.24}
{\cal F}_p=\int\frac{\mathrm d\bar k_s}{2\pi}&\sum_{\alpha,\beta\in I}
\Big\{ p_\beta p^*_\alpha c^*_\beta c_\alpha
g(\bar k_s,\Delta\tau_\beta,\Delta\tau_\alpha)\nn\\
&\times\exp[-\I(\bar k_s-\bar k_m)(\Delta\tau_\beta-\Delta\tau_\alpha)]\nn\\
&\times U^*_\beta(z_a,z_0;\bar k_m)U_\alpha(z_a,z_0;\bar k_m)\Big\}\ .
\end{align}
In (\ref{4.23}), (\ref{4.24}) and in the following only the leading terms for large $\bar k_m$ are kept. In analogy to (\ref{3.18}) we define, with $\bar k=\bar k_m$, the shift of the reduced arrival times at $z=z_a$ to be
\begin{align}\label{4.25}
\Delta\tau_\alpha={\frak{Z}}_\alpha(z_a)-(z_a-z_0)
=\frac{m}{\bar k^2_m}\int^{z_a}_{z_0}\mathrm dz'\,V_\alpha(z')
\end{align}
with $\mathcal O(\bar k_m^{-3})$ terms neglected, see (\ref{C.5}) and (\ref{C.5a}). Furthermore, we introduce the function
\begin{align}\label{4.26}
& g(\bar k_s,\Delta\tau_\beta,\Delta\tau_\alpha)
=\int\frac{\mathrm d\bar k_d}{2\pi}f^*(\bar k_s-\frac12\bar k_d)
f(\bar k_s+\frac{1}{2}\bar k_d)\nn\\
&\times\int \mathrm dz'\int \mathrm d^2x_T\exp\left[\I\bar k_d\left(z'+\frac12\Delta\tau_\beta+\frac12\Delta\tau_\alpha\right)\right]\nn\\
&\qquad\qquad\times\chi^*(\vec x_T,z'+\Delta\tau_\beta)\chi(\vec x_T,z'+\Delta\tau_\alpha)\ .
\end{align}
From (\ref{3.22}) we get with $\bar k=\bar k_m$ and $k(z_0)=k(z_a)=\bar k_m$,
\begin{align}\label{4.27}
U_\alpha(z_a,z_0;\bar k_m)=
\exp[-\I\phi_{\mathrm{dyn},\alpha}+\I\phi_{\mathrm{geom},\alpha}]
\end{align}
where the complex dynamical and geometrical phases are
\begin{align}\label{4.28}
\phi_{\mathrm{dyn},\alpha}&=\frac{m}{\bar k_m}\int^{z_a}_{z_0}\mathrm dz'[V_\alpha(z')
-\frac{\I}{2}\Gamma_\alpha(z')]\ ,\\
\phi_{\mathrm{geom},\alpha}&=\gamma_{\alpha\alpha}(z_a)-\gamma_{\alpha\alpha}(z_0)\label{4.29}\ .
\end{align}
In (\ref{4.28}) terms of $\mathcal O(\bar k_m^{-2})$ are neglected. From (\ref{4.25}) and (\ref{4.28}) we find
\begin{align}\label{4.30}
\mRe\,\phi_{\mathrm{dyn},\alpha}=\bar k_m\Delta\tau_\alpha\ .
\end{align}

In (\ref{4.24})--(\ref{4.30}) we have given our general result for the integrated flux ${\cal F}_p$ of the atom in the state $\rket{p}$ where we supposed large $\bar k_m$, see Appendix \ref{s:AppendixC}. That is, (\ref{4.24})--(\ref{4.30}) are valid for all wave packets of the type described in $\mbox{(\ref{4.2})--(\ref{4.7})}$. We insert now the Gaussian distributions for $\chi$ and $f$, see (\ref{4.8})--(\ref{4.12}). After a simple calculation presented in Appendix \ref{s:AppendixC} we obtain for this case for $\mathcal F_p$, an easily accessible observable for lABSE experiments, the result:
\begin{align}\label{4.31}
{\cal F}_p=\sum_{\alpha,\beta\in I}&p_\beta p^*_\alpha c^*_\beta c_\alpha
\exp [-(\Delta\tau_\beta-\Delta\tau_\alpha)^2/(8\sigma'^2_k)]\nn\\
&\quad\times U^*_\beta(z_a,z_0;\bar k_m)U_\alpha(z_a,z_0;\bar k_m)\ .
\end{align}

\begin{figure}[htb]
\centering
\includegraphics[scale=1.1]{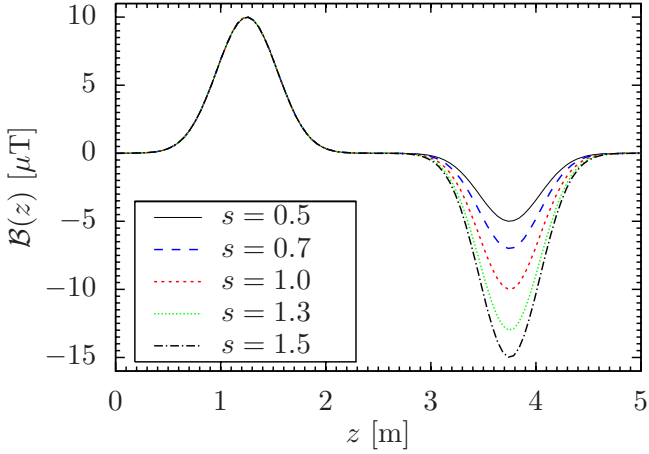}
\caption{The magnetic field as used for the calculations in this section is shown for several values of the parameter $s$ from (\ref{4.32}).}
\label{5MeterBFeld_s}
\end{figure}

\begin{figure}[htb]
\centering
\includegraphics[scale=1.1]{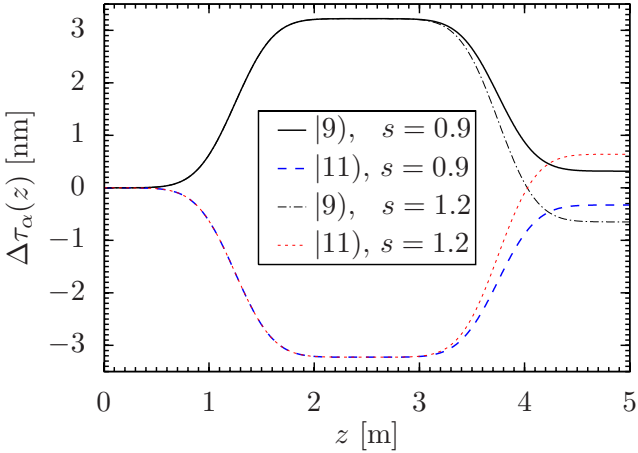}
\caption{Two examples of the roadmap plots for the states $\alpha=9$ and $\alpha=11$ based on the magnetic field (\ref{4.32}) and $\vec{\mathcal E}(z)=\vec 0$ with fixed parameters $s=0.9$ and $s=1.2$ respectively.}
\label{5MeterTimeTable_s}
\end{figure}

As a concrete application we discuss our result (\ref{4.31}) for the following field configuration. We choose the electric field as $\vec{\mathcal E}(z)=\vec 0$ and the magnetic field similar to (\ref{3.18a})
\begin{align}\label{4.32}
\vec{\mathcal B}(z)&=\mathcal B(z)\vec e_3\ ,\nn\\
\mathcal B(z)&=10\,\big(e^{-6(z-1.25)^2}-s\cdot e^{-6(z-3.75)^2}\big)\,\mu\mathrm{T}\ ,
\end{align}
where a parameter $s$ is inserted in order to detune the second half of the magnetic field configuration around the symmetric arrangement, see Figure \ref{5MeterBFeld_s}. The reason for the introduction of $s$ will be discussed below. Note that such magnetic fields can be produced by suitably shaped coils. The so-called spin echo coil produces the scalable part of $\vec{\mathcal B}(z)$ via controlling the electric current through the coil.
Figure \ref{5MeterTimeTable_s} illustrates the global behaviour of the two states $\alpha=9$ and $\alpha =11$ for different values of $s$. Setting $s=1$ leads to the roadmap plot shown in Figure \ref{TimeTable}. Considering $s=0.9$, the state $\alpha=9$ accumulates a delay compared to a free wave, while the state $\alpha=11$ picks up an advance. The opposite is true for $s=1.2$.

\begin{figure}[htb]
\centering
\includegraphics[scale=1.08]{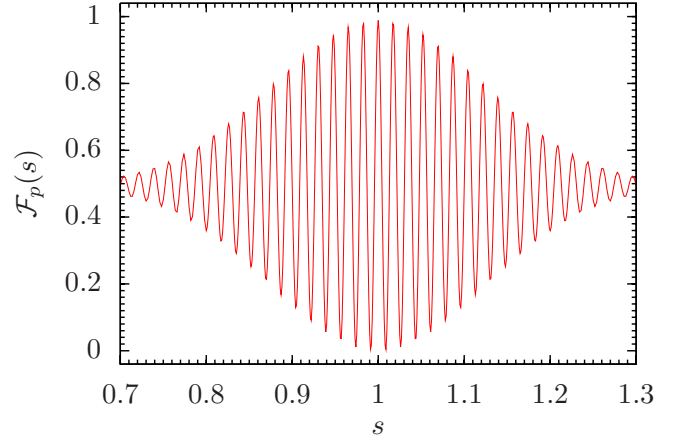}
\caption{The total integrated flux for the magnetic field in (\ref{4.32}) with the parameter $s$ varied over the interval $[0.7,1.3]$. Every value of $s$ leads to a fixed flux of analysing states.}
\label{TIF_s}
\end{figure}

Eventually, the evaluation of the total integrated flux $\mathcal F_p$ is shown in Figure \ref{TIF_s}. In our calculation we choose for the initial wave packet Gaussian momentum distributions for the functions in (\ref{4.8})--(\ref{4.11}) with
\begin{align}
\bar k_m &= 5.6\times10^{10}\,\mathrm{m}^{-1}\ ,\label{4.32b}\\
\sigma_L &=\sigma_T = 100\,\mu\mathrm{m}\ ,\label{4.32c}\\
\sigma'_k &= 0.4\,\mathrm{nm}\ .\label{4.32d}
\end{align}
Furthermore, we choose the coefficients $c_\alpha$ of the initial state the same as those of the analysing state, namely
\begin{align}\label{4.32e}
c_9=c_{11}=p_9=p_{11}=\frac{1}{\sqrt{2}}\ .
\end{align}
The initial state is then according to (\ref{4.2}) and (\ref{4.7})
\begin{align}\label{4.32f}
\left.\ket{\Psi(\vec x,t)}\right|_{t=0}&=
\Psi(\vec x,0)\frac{1}{\sqrt{2}}\left.\big(\rket{9}+\rket{11}\big)\right|_{\vec{\mathcal E}=\vec{\mathcal B}=\vec 0}\ ,
\end{align}
\begin{align}\label{4.32g}
\Psi(\vec x,0)&=\int\frac{\mathrm d^2q_T\mathrm d q_L}{(2\pi)^3}\exp[\I\vec q_T\vec x_T+\I q_L(z-z_0)]\,\tilde{\Psi}(\vec q_T,q_L)\nn\\
&=(2\pi)^{-3/4}\sigma^{-1}_T(\sigma'_k)^{-1/2}\,\exp[\I\bar k_m(z-z_0)]\nn\\
&\times\exp\left[-\frac{1}{4\sigma^2_T}\vec x^2_T-\frac{1}{4\sigma'^2_k}(z-z_0)^2\right]\ .
\end{align}
Thus, with the choice (\ref{4.32e}) we superpose the states $\alpha=9$ and $\alpha=11$ at $z_0=0$ and project the evolved state at $z_a=5\,$m onto the same superposition.

The signal to be observed experimentally is the integrated flux as function of the parameter $s$ at a fixed point on the z-axis, in our case at the final point $z_a=5\,$m of the interferometer, see Figure \ref{TIF_s}. A variation of $s$ alters the terms $(\Delta\tau_\beta-\Delta\tau_\alpha)$ in (\ref{4.31}) as well as the dynamical phases $\phi_{\mathrm{dyn}}$ in (\ref{4.28}). Thus, with the simple field configuration of Figure \ref{5MeterBFeld_s} it should be possible to create an interference signal simply by tuning the current of the spin echo coil and counting outgoing atoms. This signal could serve as a reference for more complicated field configurations.

\begin{figure}[htb]
\centering
\includegraphics[scale=0.97]{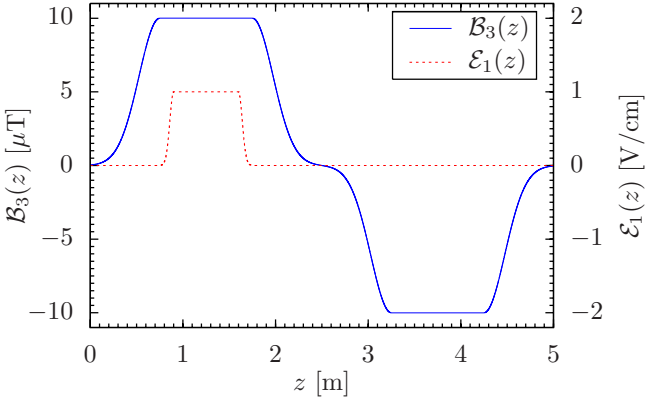}
\caption{The magnetic field component $\mathcal B_3(z)$ (\ref{advancedB}), solid line, and the electric field component $\mathcal E_1(z)$ (\ref{advancedE}), given as the dashed line.}
\label{5MeterFelder}
\end{figure}

Let us now consider a configuration with magnetic and electric fields as a second example. We define the functions
\begin{align}
\mathcal B(z)&=10\,\big(e^{-10(z - 0.75)^2}+e^{-10(z - 1.75)^2}\nn\\
&\qquad\qquad-e^{-10(z - 3.25)^2} - e^{-10(z - 4.25)^2}\big)
\end{align}
and
\begin{align}
\mathcal E(z)=e^{-300(z-0.9)^2} + e^{-300(z-1.6)^2}
\end{align}
where $z$ is given in meters. Now we choose the magnetic field $\vec{\mathcal B(z)}$, given in $\mu\mathrm{T}$, and the electric field $\vec{\mathcal E(z)}$, given in $\mathrm{V/cm}$, as follows
\begin{align}\label{advancedB}
\begin{split}
\mathcal B_1 &= \mathcal B_2 = 0\ ,\\
\mathcal B_3 &= \left\{\begin{array}{rl} \mathcal B(z) & ,\;z\in [0,0.75]\cup[1.75,2.5]\, ,\\ \mathcal B(0.75) &,\;z\in [0.75,1.75]\, ,\\ s\cdot\mathcal B(z) & ,\;z\in [2.5,3.25]\cup[4.25,5]\, ,\\ s\cdot\mathcal B(3.25) &,\; z\in [3.25,4.25]\, , \end{array}\right.\
\end{split}
\end{align}
and
\begin{align}\label{advancedE}
\begin{split}
\mathcal E_1 &= \left\{\begin{array}{rl} \mathcal E(z) & ,\;z\in [0,0.9]\cup[1.6,5]\, ,\\ \mathcal E(0.9) &,\; z\in [0.9,1.6]\, , \end{array}\right.\\
\mathcal E_2 &= \mathcal E_3=0\ ,
\end{split}
\end{align}
respectively. In Figure \ref{5MeterFelder} we show $\mathcal B_3(z)$ and $\mathcal E_1(z)$. The electric field is ramped to a constant value and back again only within a region of constant magnetic field. This procedure implies vanishing geometric phases.

\begin{figure}[htb]
\centering
\includegraphics[scale=1.1]{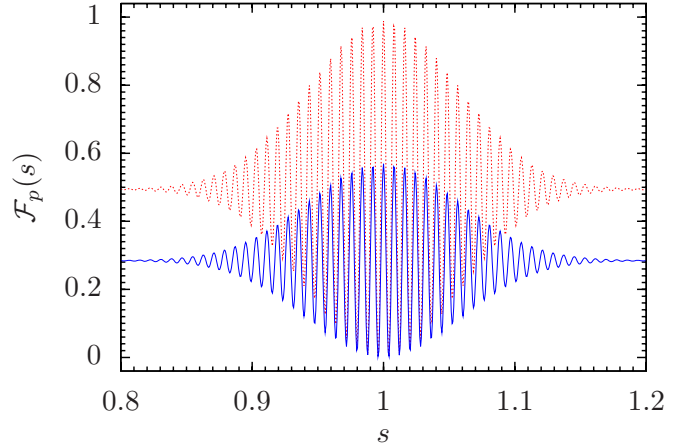}
\caption{The spin echo signal corresponding to the case where only the magnetic field (\ref{advancedB}) is present is given by the (red) dashed line. The spin echo signal corresponding to the presence of the magnetic field (\ref{advancedB}) and the electric field (\ref{advancedE}) is given by the solid line.}
\label{Spinecho_all_script}
\end{figure}

For the calculation of the spin echo signal with the external fields (\ref{advancedB}) and (\ref{advancedE}), we again employ the initial specifications (\ref{4.32b})--(\ref{4.32g}). The results for both the pure magnetic field and the combined magnetic and electric fields are shown in Figure \ref{Spinecho_all_script}. In the presence of the electric field $\vec{\mathcal E}$, we find a decreased total integrated flux $\mathcal F^{\vec{\mathcal E}\not=\vec 0}_p(s)$ due to the enhanced decay rate, compared to the case of $\mathcal F^{\vec{\mathcal E}=\vec 0}_p(s)$.

In order to interpret Figure \ref{Spinecho_all_script}, we now give an estimate of the quantities which lead to the altered spin echo signal when the electric field (\ref{advancedE}) is switched on. For $\vec{\mathcal E}\not= 0$ the quadratic Stark effect leads to a shift of the complex energies $E_\alpha$. The shift of the real parts $V_\alpha$ of $E_\alpha$ turns out to be negligible. Therefore, the reduced arrival times do not change substantially compared to the case of $\vec{\mathcal E}= 0$. The change of the oscillation frequency of $\mathcal F^{\vec{\mathcal E}\not=\vec 0}_p(s)$, induced by $V_\alpha$, is also not significant if we compare to $\mathcal F^{\vec{\mathcal E}=\vec 0}_p(s)$, see Figure \ref{Spinecho_all_script}. However, the altered imaginary parts $-\frac12\Gamma_\alpha$ have a large impact on the total integrated flux via the dynamical phases. For the case of the non-zero electric field (\ref{advancedE}), we get for the states $\alpha=9$, $10$ and $11$ the relation
\begin{align}
\Gamma^{\vec{\mathcal E}\not=\vec 0}_\alpha(z)&\approx300\, \Gamma^{(\vec{\mathcal E}= \vec 0)}_\alpha(z)\nn\\
&\approx 300\,\Gamma_S\approx2470\,\mathrm{s}^{-1}
\end{align}
for the parameter range $0.85\lesssim z\lesssim 1.65$, where $\Gamma_S$ is given in (\ref{2.18}). Taking into account that in this case the geometric phases are zero and considering (\ref{4.31}) together with (\ref{4.27}) and (\ref{4.28}), we find for $s=1$:
\begin{align}
\mathcal F^{(\vec{\mathcal E}\not= \vec 0)}_p(s=1)&\approx\exp\big(-\frac{m}{k_m}\, 300\int^{z=1.65\,\mathrm{m}}_{z=0.85\,\mathrm{m}}\mathrm{d}z'\Gamma_S\big)\nn\\
&\qquad\times\mathcal F^{(\vec{\mathcal E}=\vec 0)}_p(s=1)\ .
\end{align}
Inserting
\begin{align}
\frac{m}{k_m}\corresp (3500\,\mathrm m / \mathrm s)^{-1}\ ,
\end{align}
we get
\begin{align}\label{estimateforEnot0}
\mathcal F^{(\vec{\mathcal E}\not= \vec 0)}_p(s=1)\approx 0.569\,\mathcal F^{(\vec{\mathcal E}= \vec 0)}_p(s=1)\approx 0.562\ .
\end{align}
The rough estimate (\ref{estimateforEnot0}) for the total integrated flux coincides satisfactorily with the numerical result
\begin{align}
\mathcal F^{(\vec{\mathcal E}\not= \vec 0)}_p(s=1)\approx 0.569\ ,
\end{align}
given as the global maximum of the spin echo signal for $\vec{\mathcal E}\not=\vec 0$ in Figure \ref{Spinecho_all_script}. With this we close our discussion of the concrete examples.

In future work we will study field configurations where geometric phases, both P-conserving and P-violating ones, occur. In principle the spin echo experiments should provide the experimental access to these geometric phases since a non-vanishing geometric phase will modify the signals like that of Figures \ref{TIF_s} and \ref{Spinecho_all_script} for which only the dynamical phases are non-zero. For example, switching on suitable electric fields, geometric phases will contribute to the total phase factors $U_\beta^*$ and $U_\alpha$ in (\ref{4.31}). Besides the changed magnitudes of the peaks, the crucial information is the displacement of their positions in the $s-\mathcal F_p(s)$ interference-diagram compared to the reference signal indicating non-zero geometric phases. Of course, a P-violating Berry phase will modify the reference signal in a slightly different way if the measurement is performed with the space reflected field configuration. In this way, the high sensitivity of lABSE experiments could possibly lead to measurements of P-conserving as well as P-violating geometric phases in metastable hydrogen.

%========================================================================================
\section{Conclusions and outlook}\label{s:Conclusions}
%========================================================================================
In this article we have discussed the evolution of the meta\-stable 2S states of atomic hydrogen within the setup of longitudinal atomic beam spin echo (lABSE) experiments. The conditions for adiabatic evolution and sufficient flux at the end of the interferometer were investigated. Using the results for atoms at rest \cite{BeGaNa07_II}, we derived the solution of the effective Schr\"odinger equation for the metastable atoms for the case that the longitudinal-kinetic energy is much larger than the trans\-verse-kinetic and the potential energies. The wave function consists of a plane wave factor, a complex phase factor containing a geometric Berry phase and the envelope amplitude of the propagating state. In Section \ref{s:InterferenceEffects} we considered the interference effects between two metastable 2S states and computed the total integrated flux $\mathcal F_p$ for a given projection at the end of the interferometer in (\ref{4.31}). This quantity $\mathcal F_p$ is conveniently accessible to experiments and encodes the information from the phase differences picked up by two metastable 2S states in the interferometer. In future work the extraction of geometric phases from such experiments will be further analysed. The main result of the present paper is to provide the suitable framework for the rigorous theoretical description of stable or metastable atoms in a longitudinal spin echo device. Here we have considered the adiabatic case and abelian geometric phases. Possible extensions of the present work would be to consider non-abelian dynamic and geometric phases as well as non adiabatic motion.

%========================================================================================
\subsection*{Acknowledgements}
%========================================================================================
We would like to thank D.~Dubbers for continuous support and for useful discussions. Thanks also go to R.~Dunford, I.~Khriplovich and the other participants of the workshop "P and T Violation at Low Energies and Related Phenomena" held in June 2008 in Heidelberg for discussions of many theoretical and experimental aspects of atomic parity violation.

%****************************************************************************************

\appendix

%\titlelabel{Appendix \thetitle}

\section*{Appendix}
%========================================================================================
\section{Values for quantities related to $n=2$ hydrogen}\label{s:AppendixA}
%========================================================================================

\renewcommand{\theequation}{\thesection.\arabic{equation}}
 \setcounter{equation}{0}

In this appendix we collect the numerical values for the quantities entering our calculations for the hydrogen states with principal quantum number $n=2$. We specify our numbering scheme for these states and give the expressions for the mass matrix at zero external fields, and for the electric and the magnetic dipole operators.

In Table \ref{t:values} we present the numerical values for the weak charges $Q_W^{(\varkappa)}$, $\varkappa=1,2$, the quantities $\Delta q$, the Lamb shift $L = E(2\mathrm S_{1/2})-E(2\mathrm P_{1/2})$, and the fine structure splitting $\Delta = E(2\mathrm P_{3/2}) - E(2\mathrm P_{1/2})$. Here $\Delta q$ denotes the total polarisation of the nucleus carried by the quark species $q$ $(q=u,d,s)$. The ground state hyperfine splitting energy is denoted by $\mcal A$. We have $\mcal A = E(1\mathrm S_{1/2},F=1)-E(1\mathrm S_{1/2},F=0)$ for hydrogen. For the precise definitions of all these quantities see \cite{BeGaNa07_II}.
The $n=2$ states of hydrogen in the absence of P-violation and for zero external fields are denoted by $\rket{2L_J,F,F_3}$, where $L$, $J$, $F$ and $F_3$ are the quantum numbers of the electron's orbital angular momentum, its total angular momentum, the total atomic angular momentum and its third component, respectively. The quantum numbers $S$ for the electron spin and $I$ for the nuclear spin are omitted, since these are fixed quantities. In the following the ordering of the atomic states in the matrix representations of operators is according to Table \ref{t:state.labels} where we give the numbering scheme for the states which we consider. For electric field $\vmc E$ and magnetic field $\vmc B$ equal to zero we have the free 2S and 2P states. We write $\hat L$, $\hat P$, $\hat S$ since these states include the parity mixing due to $H_{\PV}$, see (\ref{II.e1:eff.HPV}) of \cite{BeGaNa07_II}.
\bgroup
\tabcolsep=0.2em
\begin{table}[Hb!]
\begin{center}
\begin{tabular}{ccc}
\hline\noalign{\smallskip}
 		 & ${}_1^1$H 					& Ref.\\   
\noalign{\smallskip}\hline\noalign{\smallskip}
$Z$ 		 & 1 						&\\
$N$ 		 & 0 						&\\
$I$ 		 & $\tfrac12$ 					&\\    
\noalign{\smallskip}\hline\noalign{\smallskip}
$Q_W^{(1)}(Z,N)$ & 0.04532(64) 					& (\ref{II.e1:QW1.SM}) of \cite{BeGaNa07_II}\\ 
$\delta_1(Z,N)$  & $-2.78(4)\cdot 10^{-13}$ 			& (\ref{II.e2:deltaPV12H}) of \cite{BeGaNa07_II}\\   
\noalign{\smallskip}\hline\noalign{\smallskip}
$\Delta u(Z,N) - \Delta d(Z,N)$ & $1.2695(29)$ 			&  \cite{PDG06}\\
$\Delta s(Z,N)$  & $0.006(29)(7)$ 				& \cite{Jac06} \\
$Q_W^{(2)}(Z,N)$ & $-0.1145(31)$ 				& (\ref{II.e1:QW2.SM}) of \cite{BeGaNa07_II}\\
$\delta_2(Z,N)$  & $7.04(19)\cdot 10^{-13}$ 			& (\ref{II.e2:deltaPV12H}) of \cite{BeGaNa07_II}\\   
\noalign{\smallskip}\hline\noalign{\smallskip}
%$L(Z,N)/h$ & \multicolumn{1}{r|}{1057.8447(34) MHz} & \multicolumn{1}{r|}{1059.2338(34) MHz} & \cite{BiHa01}\\
$L(Z,N)/h$ 	 & \multicolumn{1}{r}{1057.8440(24) MHz} 	& \cite{NISTData}\\
$\Delta(Z,N)/h$  & \multicolumn{1}{r}{10969.0416(48) MHz} 	& \cite{NISTData}\\
$\mcal A(Z,N)/h$ & \multicolumn{1}{r}{\ 1420.405751768(1) MHz} 	& \cite{Kar05}\\   
\noalign{\smallskip}\hline\noalign{\smallskip}
\end{tabular}
\caption{Values of parameters for numerical calculations. The weak mixing angle in the low energy limit, $\sin^2\vartheta_W = 0.23867(16)$, was taken from \cite{ErRa05}. The uncertainty in $\delta_1$ is dominated by the uncertainty of $\sin^2\vartheta_W$. The uncertainties in $Q_W^{(2)}$ and $\delta_2$ for hydrogen are resulting from the errors of the weak mixing angle and $\Delta s$ in equal shares.}
\label{t:values}
\end{center}
\end{table}
\egroup
Consider first atoms in a constant $\vmc B$-field pointing in positive 3-direction,
\begin{align}\label{eA:B3}
\vmc B = \mcal B\vec e_3\ ,\quad \mcal B> 0\ .
\end{align}
The corresponding states $\rket{2\hat L_J, F,F_3,0,\mcal B\vec e_3}$ are obtained from those at $\mcal B=0$ by continuously turning on $\vmc B$ in the form (\ref{eA:B3}). Of course, for $|\vmc B| \neq 0$, $F$ is no longer a good quantum number. Here it is merely a label for the states.
We now choose a reference field $\vmc B_\mathrm{ref}=\mcal B_\mathrm{ref}\vec e_3$, $\mcal B_\mathrm{ref}>0$, below the first crossings in the Breit-Rabi diagrams, for instance $\mcal B_{\mathrm{ref}} = 0.05\,\mathrm{mT}$. We define the states $\rket{2\hat L_J,F,F_3,$ $\vmc E,\vmc B}$ for arbitrary $\vmc B$ fields in the neighbourhood of $\vmc B_\mathrm{ref}$ as the states obtained continuously from $\rket{2\hat L_J,F,F_3,0,\vmc B_\mathrm{ref}}$ by turning on $\vmc E$ and $(\vmc B-\vmc B_\mathrm{ref})$ as $\lambda\vmc E$ and $\lambda(\vmc B-\vmc B_\mathrm{ref})$, respectively, with $\lambda\in[0,1]$. Then both $F$ and $F_3$ are only labels for these states and no longer good quantum numbers.

\newcommand{\phm}{\phantom-}
{%\squeezetable

\begin{table}[htbp]
\begin{center}
\begin{tabular}{cl}
\cline{1-2}\noalign{\smallskip}
\multicolumn{2}{c}{hydrogen}\\ 
\noalign{\smallskip}\cline{1-2}\noalign{\smallskip}
$\alpha$ & $\rket{2\hat L_J,F,F_3,\vmc E,\vmc B}$\\ 
\noalign{\smallskip}\cline{1-2}\noalign{\smallskip}
1 & $\rket{2\hat P_{3/2},2,\phm2,\vmc E,\vmc B}$\\
2 & $\rket{2\hat P_{3/2},2,\phm1,\vmc E,\vmc B}$\\
3 & $\rket{2\hat P_{3/2},2,\phm0,\vmc E,\vmc B}$\\
4 & $\rket{2\hat P_{3/2},2,-1,\vmc E,\vmc B}$   \\
5 & $\rket{2\hat P_{3/2},2,-2,\vmc E,\vmc B}$   \\
6 & $\rket{2\hat P_{3/2},1,\phm1,\vmc E,\vmc B}$\\
7 & $\rket{2\hat P_{3/2},1,\phm0,\vmc E,\vmc B}$\\
8 & $\rket{2\hat P_{3/2},1,-1,\vmc E,\vmc B}$   \\ 
\noalign{\smallskip}\cline{1-2}\noalign{\smallskip}
9 & $\rket{2\hat S_{1/2},1,\phm1,\vmc E,\vmc B}$\\
10 & $\rket{2\hat S_{1/2},1,\phm0,\vmc E,\vmc B}$\\
11 & $\rket{2\hat S_{1/2},1,-1,\vmc E,\vmc B}$   \\
12 & $\rket{2\hat S_{1/2},0,\phm0,\vmc E,\vmc B}$\\
\noalign{\smallskip}\cline{1-2}\noalign{\smallskip}
13 & $\rket{2\hat P_{1/2},1,\phm1,\vmc E,\vmc B}$\\ 
14 & $\rket{2\hat P_{1/2},1,\phm0,\vmc E,\vmc B}$\\ 
15 & $\rket{2\hat P_{1/2},1,-1,\vmc E,\vmc B}$   \\ 
16 & $\rket{2\hat P_{1/2},0,\phm0,\vmc E,\vmc B}$\\ 
\noalign{\smallskip}\cline{1-2} 
\end{tabular}
\caption{The numbering scheme for the atomic states of hydrogen.}
\label{t:state.labels}\vskip10pt
\end{center}
\end{table}
Tables \ref{t:H2.M0}--\ref{t:H2.Mu} in Appendix \ref{s:AppendixD} show the non-zero parts of the mass matrix $\umat{\tilde M}_0(\delta_1,\delta_2)$ of (\ref{2.4b}) for hydrogen for zero external fields, of the electric dipole operator $\uvec{D}$ and of the magnetic dipole operator $\uvec{\mu}$ for the $n=2$ states.

Next we discuss the conditions for decoupling of the evolution of the 2S states from the 2P states. We use for that the results of Appendix \ref{s:AppendixB} of \cite{BeGaNa07_II}. Let
\begin{align}\label{A.100}
T\approx 1\, \mathrm{ms}
\end{align}
be the total observation time. We then have, with $v_z$ from (\ref{2.16}), $v_z T\approx 3.5\,\mathrm m$ which is of the same size as $l$ (\ref{2.15}). From (\ref{II.B.1})--(\ref{II.B.3}) of \cite{BeGaNa07_II} we get as conditions for the decoupling
\begin{align}\label{A.101}
\frac{2c_{12}}{T\Delta\Gamma_{\mathrm{min}}}&\ll 1\ ,\\
\frac{4c_{12}c_{21}}{T\Delta\Gamma_{\mathrm{min}}}&\ll 1\ ,\label{A.101a}\\
\frac{2c_{21}}{T\Delta\Gamma_{\mathrm{min}}}&\ll 1 \label{A.101b}
\end{align}
with
\begin{align}\label{A.102}
\Delta\Gamma_{\mathrm{min}}\approx\Gamma_P=\tau^{-1}_P\ .
\end{align}
In the following we shall use with $\tau_P$ from (\ref{2.18})
\begin{align}\label{A.102a}
v_zT&\approx l\approx5\,\mathrm m\ ,\nn\\
v_z\tau_P&=5.6\,\mathrm{\mu m}\ .
\end{align}
As in \cite{BeGaNa07_II} we shall use for our estimates the quantities ${\cal E}_0$ and ${\cal B}_0$ which characterise the values of the field strengths where the electric and magnetic energy, respectively, reach the value of the Lamb shift. With $r_B(1)$ the Bohr radius of the H atom and $\mu_B$ the Bohr magneton we have (see (\ref{II.B.7}), (\ref{II.B.16}) and (\ref{II.B.16a}) of \cite{BeGaNa07_II}
\begin{align}\label{A.103}
{\cal E}_0&=\frac{L}{\sqrt{3}er_B(1)}=477.3\,\mathrm{V/cm}\ ,\\
{\cal B}_0&=\frac{L}{\sqrt{3}\mu_{\cal B}}=43.65 \,\mathrm{mT}\ .\label{A.104}
\end{align}
For a time-varying electric field $\vec{\mathcal E}(t)$ the quantities $c_{12}$ and $c_{21}$ are estimated in (\ref{II.B.4})--(\ref{II.B.6}) of \cite{BeGaNa07_II}. We take this over for our case with the replacement $\mathrm dt\to v^{-1}_z\mathrm dz$, that is,
\begin{align}\label{A.105}
\left|\frac{\partial\vec{\mathcal E}(t)}{\partial t}\right|\to
v_z\left|\frac{\partial\vec{\mathcal E}(z)}{\partial z}\right|\ .
\end{align}
In this way we get
\begin{align}\label{A.106}
c_{12}\approx c_{21}\approx
\frac{Tv_z}{\sqrt{3}{\mathcal E_0}}\max
\left|\frac{\partial\vec{\mathcal E}(z)}{\partial z}\right|\ .
\end{align}
Here and in the following the maximum is to be taken for $z_0\leq z\leq z_a$. We set
\begin{align}\label{A.107}
\max\left|\frac{\partial\vec{\mathcal E}(z)}{\partial z}\right|
=\frac{\Delta{\cal E}}{\Delta z}
\end{align}
with a typical variation (see (\ref{2.20}))
\begin{align}\label{A.108}
\Delta{\cal E}\approx 1\mathrm{V/cm}
\end{align}
and $\Delta z$ the allowed variation length which we want to estimate. Inserting (\ref{A.107}) in (\ref{A.106}) we get from (\ref{A.101}) and (\ref{A.101b})
\begin{align}\label{A.109}
\Delta z\gg\frac{2}{\sqrt{3}}v_z\tau_P\frac{\Delta{\cal E}}{{\cal E}_0}
\approx 0.02\,\mathrm{\mu m}\ .
\end{align}
From (\ref{A.101a}) we get
\begin{align}\label{A.110}
\Delta z\gg\frac{2}{\sqrt{3}}\sqrt{v_zTv_z\tau_P}\frac{\Delta{\cal E}}{{\cal E}_0}
\approx 12\,\mathrm{\mu m}\ .
\end{align}
Thus, for electric field variations (\ref{A.107}), (\ref{A.108}) a length scale
\begin{align}\label{A.111}
\Delta z\geq 50\,\mathrm{\mu m}
\end{align}
should be sufficient to guarantee decoupling of the 2S from the 2P states.

The allowed time variation of magnetic fields was estimated in (\ref{II.B.14})--(\ref{II.B.18}) of \cite{BeGaNa07_II} but allowing rather large electric fields of order ${\cal E}_0/2$. Here we allow only fields of order
\begin{align}\label{A.112}
{\cal E}=1\,\mathrm{V/cm}\ .
\end{align}
Thus we get from (\ref{II.B.4}) and (\ref{II.B.15}) of \cite{BeGaNa07_II} with the insertion of an additional factor $2{\cal E}/{\cal E}_0$ and the replacement
\begin{align}\label{A.113}
\left|\frac{\partial\vec{\mathcal B}(t)}{\partial t}\right|\to v_z
\left|\frac{\partial\vec{\mathcal B}(z)}{\partial z}\right|
\end{align}
the estimate
\begin{align}\label{A.114}
c_{12}\approx c_{21}\approx\frac{2}{\sqrt{3}}Tv_z\frac{{\mathcal E}}{{\mathcal E}_0{\mathcal B}_0}\max\left|\frac{\partial{\vec{\mathcal B}}(z)}{\partial z}\right|\ .
\end{align}
We set here
\begin{align}\label{A.115}
\max\left|\frac{\partial{\vec{\mathcal B}}(z)}{\partial z}\right|&=
\frac{\Delta {\cal B}}{\Delta z}\ ,\nn\\
\Delta {\cal B}&=1 \,\mathrm{mT}\ .
\end{align}
Inserting now (\ref{A.114}) in (\ref{A.101}) and (\ref{A.101b}) we get
\begin{align}\label{A.116}
\Delta z\gg
\frac{4}{\sqrt{3}}v_z\tau_P\frac{{\cal E}\Delta{\cal B}}{{\cal E}_0{\cal B}_0}
\approx 6\times 10^{-4}\,\mathrm{\mu m}\ .
\end{align}
From (\ref{A.101a}) we get
\begin{align}\label{A.117}
\Delta z\gg\frac{4}{\sqrt{3}}
\sqrt{v_zTv_z\tau_P}\frac{{\cal E}\Delta {\cal B}}{{\cal E}_0{\cal B}_0}
\approx 0.6\,\mathrm{\mu m}\ .
\end{align}
Thus, a variation length $\Delta z\approx 10\,\mathrm{\mu m}$ should be sufficient to ensure decoupling of the 2S from the 2P states for the magnetic fields considered. 

%========================================================================================
\section{Solution for one component, details}\label{s:AppendixB}
%========================================================================================
 \setcounter{equation}{0}

In this appendix we give the detailed derivation of the results of Section \ref{s:SolutionOneComp}. First we derive the invariance of $\mathcal V_\alpha(z)$ in (\ref{2.35}) under the local phase transformations (\ref{2.35b}) where the functions $\eta_\alpha(z)$ are assumed to be real and differentiable.
From (\ref{2.32}) we obtain
\begin{align}\label{B.0}
\frac{\partial}{\partial z}\gamma_{\alpha\alpha}(z)=\I\lrbra{\alpha(z)}\frac{\partial}{\partial z}\rket{\alpha(z)}\ .
\end{align}
Hence the transformation (\ref{2.35b}) yields
\begin{align}\label{B.0a}
\left(\frac{\partial}{\partial z}\gamma_{\alpha\alpha}(z)\right)^2\longrightarrow\left(\I\lrbra{\alpha(z)}\frac{\partial}{\partial z}\rket{\alpha(z)}-\frac{\partial}{\partial z}\eta_\alpha(z)\right)^2
\end{align}
and
\begin{align}\label{B.0b}
\left(\frac{\partial}{\partial z}\lrbra{\alpha(z)}\right)\frac{\partial}{\partial z}\rket{\alpha(z)}&\longrightarrow\left(\frac{\partial}{\partial z}\eta_\alpha(z)\right)^2\nn\\
&\quad-2\I\left(\frac{\partial}{\partial z}\eta_\alpha(z)\right)\lrbra{\alpha(z)}\frac{\partial}{\partial z}\rket{\alpha(z)}\nn\\
&\quad +\left(\frac{\partial}{\partial z}\lrbra{\alpha(z)}\right)\frac{\partial}{\partial z}\rket{\alpha(z)}\ .
\end{align}
This implies invariance of the potential $\mathcal V_\alpha(z)$:
\begin{align}\label{B.0c}
\mathcal V_\alpha(z)&\longrightarrow\mathcal V'_\alpha(z) = E_\alpha(z)+E_0\nn\\
&\quad -\frac{1}{2m}\Bigg[-\left(\lrbra{\alpha(z)}\frac{\partial}{\partial z}\rket{\alpha(z)}\right)^2\nn\\
&\quad -2\I\left(\frac{\partial}{\partial z}\eta_\alpha(z)\right)\lrbra{\alpha(z)}\frac{\partial}{\partial z}\rket{\alpha(z)}+\left(\frac{\partial}{\partial z}\eta_\alpha(z)\right)^2\nn\\
&\quad -\left(\frac{\partial}{\partial z}\eta_\alpha(z)\right)^2+2\I\left(\frac{\partial}{\partial z}\eta_\alpha(z)\right)\lrbra{\alpha(z)}\frac{\partial}{\partial z}\rket{\alpha(z)}\nn\\
&\quad -\left(\frac{\partial}{\partial z}\lrbra{\alpha(z)}\right)^{\left.\right.}\frac{\partial}{\partial z}\rket{\alpha(z)}\Bigg]\nn\\
&=\mathcal V_\alpha(z)\ .
\end{align}

Next we discuss the derivation of (\ref{3.9}), the equation for the envelope function $\mathcal A$, and of the solution (\ref{3.10}). We use the variables $\tau,\zeta$ defined in (\ref{3.3}) and make the ansatz (\ref{3.4}) for the wave function with, in general, complex $\phi(z,t)$. Inserting (\ref{3.4}) in (\ref{3.1}) gives
\begin{align}\label{B.1}
2\I\left(m\frac{\partial\mathcal A}{\partial t}+\frac{\partial\phi}{\partial z}\frac{\partial\mathcal A}{\partial z}\right)&+\Bigg[-2m\frac{\partial\phi}{\partial t}-2mV(z)+\I m\Gamma(z)\nn\\
&\hspace{1.8em}-\left(\frac{\partial\phi}{\partial z}\right)^2+\I\frac{\partial^2\phi}{\partial z^2}\Bigg]\mathcal A\nn\\
&+ \Delta_T\mathcal A + \frac{\partial^2\mathcal A}{\partial z^2} = 0\ .
\end{align}
The next step is to insert $\phi$ from (\ref{3.7}) and to change variables from $\vec x_T,z,t$ to $\vec x_T,\zeta,\tau$, see (\ref{3.3}), (\ref{3.3a}). Some useful relations in this context, with $k(z)$ defined in (\ref{3.6}), are
\begin{align}\label{B.2}
\frac{\partial}{\partial t}&=\frac{\bar k}{m}\frac{\partial}{\partial \tau}\ ,\nn\\
\frac{\partial}{\partial z}&=\frac{\bar k}{k(z)}\frac{\partial}{\partial \zeta}\ ,
\end{align}
\begin{align}\label{B.3}
\frac{\partial k(z)}{\partial z}=-\frac{m}{k(z)}
\frac{\partial V(z)}{\partial z}\ .
\end{align}
Furthermore we have
\begin{align}\label{B.4}
\mathrm d\tau&=\frac{\bar k}{m}\mathrm dt\ ,\nn\\
\mathrm d\zeta&=\frac{\bar k}{k(z)}\mathrm dz\ .
\end{align}
The local longitudinal velocity of the atom at position $z$ is
\begin{align}\label{B.5}
v(z)=\frac{k(z)}{m}\ .
\end{align}

Suppose now that the atom arrives at position $z$ at time $t$. Then it has to start at a certain position $u(z,t)$ at time $t=0$. We have
\begin{align}\label{B.6}
\mathrm dz&=v(z)\mathrm dt,\\
\int^z_{u(z,t)}\frac{\mathrm dz'}{v(z')}&=\int^t_0\mathrm dt'=t\ ,\nn\\
\int^z_{u(z,t)}\mathrm dz'\frac{\bar k}{k(z')}&=\frac{\bar k}{m}t=\tau\ .\label{A.7}
\end{align}
We can rewrite this as
\begin{align}\label{B.7a}
\int^z_{z_0}\mathrm dz'\frac{\bar k}{k(z')}
-\int^{u(z,t)}_{z_0}\mathrm dz'\frac{\bar k}{k(z')}&=\tau\\
\Rightarrow \quad\int^{u(z,t)}_{z_0}\mathrm dz'\frac{\bar k}{k(z')}&=\zeta-\tau\ .\label{B.7b}
\end{align}
Thus $u(z,t)$ is a function of $\zeta-\tau$ only. We can express this with the help of the function $\mathcal Z(\cdot)$ defined in (\ref{3.3a}):
\begin{align}\label{B.7c}
u(z,t)=\mathcal Z(\zeta-\tau)=\mathcal Z(\frak{Z}(z)-\frac{\bar k}{m}t)\ .
\end{align}
Expressed in the $\tau,\zeta$ coordinates we have the following. The atom arrives at reduced time $\tau$ at position $\zeta$ if it starts at $\tau=0$ at position $\zeta-\tau$. From (\ref{A.7}) we get easily
\begin{align}\label{B.11}
\frac{\partial u(z,t)}{\partial t}&=-\frac{k(u(z,t))}{m}\ ,\nn\\
\frac{\partial u(z,t)}{\partial z}&=\frac{k(u(z,t))}{k(z)}\ .
\end{align}
Using this we find for the partial derivatives of $\phi$ (\ref{3.7}) the following:
\begin{align}\label{B.10}
\frac{\partial\phi(z,t)}{\partial t}=-\frac{\bar k^2}{2m}-\frac{\I}{2}\frac{1}{k(u)}\frac{\partial V(u)}{\partial u}+\frac{\I}{2}\Gamma(u)\ .
\end{align}
Here and in the sequel $k(u)$ is to be read as $k(u(z,t)),~\Gamma(u)$ as $\Gamma(u(z,t))$, etc. 
\begin{align}\label{B.12}
\frac{\partial \phi(z,t)}{\partial z}&=k(z)
-\frac{\I m}{2k(z)}\bigg\{\frac{1}{k(z)}\frac{\partial V(z)}{\partial z}-\Gamma(z)\nn\\
&\hspace{8em}-(z\to u)\bigg\}\ ,\\
\frac{\partial^2\phi(z,t)}{\partial z^2}&=-\frac{m}{k(z)}\frac{\partial V(z)}{\partial z}\nn\\
&-\I\frac{m^2}{2k^3(z)}\frac{\partial V(z)}{\partial z}\bigg[\frac{1}{k(z)}\frac{\partial V(z)}{\partial z}-\Gamma(z)\nn\\
&\hspace{8.5em}-(z\to u)\bigg]\nn\\
&-\I\frac{m}{2k^2(z)}\Bigg[\frac{m}{k^2(z)}\left(\frac{\partial V(z)}{\partial z}\right)^2\nn\\
&+\frac{\partial^2 V(z)}{\partial z^2}-k(z)\frac{\partial \Gamma(z)}{\partial z}-(z\to u)\Bigg]\ .\label{B.13}
\end{align}

Inserting (\ref{B.11})--(\ref{B.13}) in (\ref{B.1}) we get the following equation for $\mathcal A(\vec x_T,\zeta,\tau)$:
\begin{align}\label{B.14}
&\left(m\frac{\partial}{\partial t}+k(z)\frac{\partial}{\partial z}\right)\mathcal A(\vec x_T,\zeta,\tau)=\nn\\
&\frac{\I}{2}\Bigg\{\frac{m^2}{4k^2(z)}
\left[\frac{1}{k(z)}\frac{\partial V(z)}{\partial z}-\Gamma(z)-(z\to u)\right]^2\nn\\
&+\frac{m^2}{2k^3(z)}\frac{\partial V(z)}{\partial z}\left[\frac{1}{k(z)}\frac{\partial V(z)}{\partial z}-\Gamma(z)-(z\to u)\right]\nn\\
&+\frac{m}{2k^2(z)}\Bigg[\frac{m}{k^2(z)}\left(\frac{\partial V(z)}{\partial z}\right)^2+\frac{\partial^2 V(z)}{\partial z^2}\nn\\
&-k(z)\frac{\partial \Gamma(z)}{\partial z}-(z\to u)\Bigg]\Bigg\}\mathcal A(\vec x_T,\zeta,\tau)\nn\\
&+\frac{\I}{2}\frac{m}{k(z)}
\left[\frac{1}{k(z)}\frac{\partial V(z)}{\partial z}-\Gamma(z)-(z\to u)\right]
\frac{\partial\mathcal A}{\partial z}(\vec x_T,\zeta,\tau)\nn\\
&+\frac{\I}{2}\Delta_T \mathcal A(\vec x_T,\zeta,\tau)+\frac{\I}{2}\frac{\partial^2}{\partial z^2}\mathcal A(\vec x_T,\zeta,\tau)\ .
\end{align}
Replacing the derivatives $\partial/\partial t$ and $\partial/\partial z$ by $\partial/\partial\tau$ and $\partial/\partial\zeta$ according to (\ref{B.2}) leads to 
\begin{align}\label{B.15}
\left(\frac{\partial}{\partial \tau}+\frac{\partial}{\partial\zeta}\right)
\mathcal A(\vec x_T,\zeta,\tau)=(L\mathcal A)(\vec x_T,\zeta,\tau)\ .
\end{align}
Here $L$ is the following differential operator
\begin{align}
L=l_0(z,u)+l_1(z,u)\frac{\partial}{\partial\zeta}+l_2(z)\frac{\partial^2}{\partial\zeta^2}+\frac{\I}{2\bar k}\Delta_T\ ,\label{B.16}
\end{align}
\begin{align}
l_0(z,u)&=\frac{\I}{2\bar k}\Bigg\{\frac{m^2}{4k^2(z)}\left[
\frac{1}{k(z)}\frac{\partial V(z)}{\partial z}-\Gamma(z)-(z\to u)\right]^2\nn\\
&+\frac{m^2}{2k^3(z)}\frac{\partial V(z)}{\partial z}
\left[\frac{1}{k(z)}\frac{\partial V(z)}{\partial z}-\Gamma(z)-(z\to u)\right]\nn\\
&+\frac{m}{2k^2(z)}\Bigg[\frac{m}{k^2(z)}\left(\frac{\partial V(z)}{\partial z}\right)^2+\frac{\partial^2 V(z)}{\partial z^2}\nn\\
&-k(z)\frac{\partial \Gamma(z)}{\partial z}-(z\to u)\Bigg]\Bigg\}\ ,\label{B.17}\\
l_1(z,u)&=\frac{\I m}{2k^2(z)}\left[\frac{1}{k(z)}\frac{\partial V(z)}{\partial z}-\Gamma(z)-(z\to u)\right]\nn\\
&+\I\frac{m}{2k^3(z)}\frac{\partial V(z)}{\partial z}\ ,\label{B.18}\\
l_2(z)&=\I\frac{\bar k}{2k^2(z)}\ .\label{B.19}
\end{align}
In (\ref{B.16})--(\ref{B.19}) we have to consider $z$ and $u$ as functions of $\zeta$ and $\tau$ according to
\begin{align}\label{B.20}
z&=\mathcal Z(\zeta)\ ,\nn\\
u&=\mathcal Z(\zeta-\tau)\ ,
\end{align}
see (\ref{3.3a}) and (\ref{B.7c}).
We are interested in the case where (\ref{2.5a}) holds which implies also 
\begin{align}\label{B.20a}
\bar k^2\gg|2mV(z)|\ .
\end{align}
Then $k(z)\approx\bar k$ and we find that all terms of $L$ have at least one power of $1/\bar k$. Thus, for large $\bar k$ we get by formal power counting
\begin{align}\label{B.20b}
L\propto\frac{1}{\bar k}\ .
\end{align}

The next step is to define the Green's function for (\ref{B.15}):
\begin{align}\label{B.21}
\left(\frac{\partial}{\partial\tau}+\frac{\partial}{\partial\zeta}\right)G(\zeta,\tau)=\delta(\zeta)\delta(\tau)\ .
\end{align}
One solution is
\begin{align}\label{B.22}
G(\zeta,\tau)=\theta(\zeta+\tau)\delta(\zeta-\tau)\ .
\end{align}
With (\ref{B.22}) we can transform (\ref{B.15}) into an integral equation
\begin{align}\label{B.23}
\mathcal A(\vec x_T,\zeta,\tau)&=B(\vec x_T,\zeta-\tau)+\int^\infty_{-\infty}\mathrm d\zeta'\int^\infty_{-\infty}\mathrm d\tau'\nn\\
&\quad\times G(\zeta-\zeta',\tau-\tau')(L\mathcal A)(\vec x_T,\zeta',\tau')\nn\\
&=B(\vec x_T,\zeta-\tau)\nn\\
&\quad+\int^\zeta_{-\infty}\mathrm d\zeta'\left.(L\mathcal A)(\vec x_T,\zeta',\tau')\right|_{\tau'=\tau-\zeta+\zeta'}\ .
\end{align}
Here $B(\vec x_T,\zeta-\tau)$ satisfies the homogeneous equation
\begin{align}\label{B.24}
\left(\frac{\partial}{\partial \tau}+\frac{\partial}{\partial \zeta}\right)B(\vec x_T,\zeta-\tau)=0
\end{align}
and is determined by the initial condition at $\tau=0$. Indeed, let us define
\begin{align}\label{B.25}
\varphi(\vec x_T,\zeta):=\left.\mathcal A(\vec x_T,\zeta,\tau)\right|_{\tau=0}\ .
\end{align}
We set
\begin{align}\label{B.26}
&B(\vec x_T,\zeta-\tau)=\varphi(\vec x_T,\zeta-\tau)\nn\\
&-\int^{\zeta-\tau}_{-\infty}\mathrm d\zeta'\left.(L\mathcal A)
(\vec x_T,\zeta',\tau')\right|_{\tau'=\tau-\zeta+\zeta'}\ .
\end{align}
and obtain from (\ref{B.23})
\begin{align}\label{B.27}
\mathcal A(\vec x_T,\zeta,\tau)&=\varphi(\vec x_T,\zeta-\tau)\nn\\
&\quad+\int^\zeta_{\zeta-\tau}\mathrm d\zeta'(L\mathcal A)
\left.(\vec x_T,\zeta',\tau')\right|_{\tau'=\tau-\zeta+\zeta'}\ .
\end{align}
We write this in matrix notation as
\begin{align}\label{B.28}
\mathcal A=\varphi+K\mathcal A\ ,
\end{align}
where we define the operator $K$ as follows: For any function $f(\vec x_T,\zeta,\tau)$ we set
\begin{align}\label{B.29}
(Kf)(\vec x_T,\zeta,\tau)
=\int^\zeta_{\zeta-\tau} \mathrm d\zeta'\left.(Lf)(\vec x_T,\zeta',\tau')\right|_{\tau'=\tau-\zeta+\zeta'}\ .
\end{align}
Since $L\propto 1/\bar k$, see (\ref{B.20b}), we have also 
\begin{align}\label{B.30}
K\propto 1/\bar k\ .
\end{align}
That is, all terms of $K$ have an explicit factor $1/\bar k$. 
The solution of (\ref{B.28}) is easily obtained
\begin{align}\label{B.31}
\mathcal A=(1-K)^{-1}\varphi = \sum^\infty_{n=0}K^n\varphi\ .
\end{align}
We write this as 
\begin{align}\label{B.32}
\mathcal A=\mathcal A^{(0)}+\mathcal A^{(1)}+\dots
\end{align}
where
\begin{align}\label{B.33}
\mathcal A^{(0)}(\vec x_T,\zeta,\tau)&=\varphi(\vec x_T,\zeta-\tau)\ ,\nn\\
\mathcal A^{(1)}(\vec x_T,\zeta,\tau)&=(K\mathcal A^{(0)})(\vec x_T,\zeta,\tau)\ ,\nn\\
\mathcal A^{(2)}(\vec x_T,\zeta,\tau)&=(K^2\mathcal A^{(0)})(\vec x_T,\zeta,\tau)\ ,\nn\\
&\hspace{0.55em}\vdots \hspace{8.6em} \ .
\end{align}
The successive terms in (\ref{B.32}) have higher and higher powers of $1/\bar k$ according to (\ref{B.30}) and are thus more and more suppressed for large $\bar k$. This is, of course, only a formal argument. With (\ref{B.32}) and (\ref{B.33}) we have given the proof of (\ref{3.9})--(\ref{3.10}). The zero-order term is further discussed in Section \ref{s:SolutionOneComp}. The range of validity of the expansion (\ref{B.32}) is discussed below.

Let us consider now the first-order term $\mathcal A^{(1)}$ in (\ref{B.33}) for a general function $\varphi$. With (\ref{B.16})--(\ref{B.20}) and (\ref{B.29}) we get
\begin{align}\label{B.34}
&(L\mathcal A^{(0)})\left.(\vec x_T,\zeta',\tau')\right|_{\tau'=\tau-\zeta+\zeta'}
=\Big\{l_0(\mathcal Z(\zeta'),\mathcal Z(\zeta'-\tau'))\nn\\
&\quad+l_1(\mathcal Z(\zeta'),\mathcal Z(\zeta'-\tau'))\frac{\partial}{\partial\zeta'}
+l_2(\mathcal Z(\zeta'))\frac{\partial^2}{\partial\zeta'^2}\nn\\
&\quad+\frac{\I}{2\bar k}\Delta_T\Big\}\left.\varphi(\vec x_T,\zeta'-\tau')
\right|_{\tau'=\tau-\zeta+\zeta'}\nn\\
&=\left\{l_0(\mathcal Z(\zeta'),\mathcal Z(\zeta-\tau))+l_1(\mathcal Z(\zeta'),\mathcal Z(\zeta-\tau))\frac{\partial}{\partial\zeta}\right.\nn\\
&\quad\left.+l_2(\mathcal Z(\zeta'))\frac{\partial^2}{\partial\zeta^2}+\frac{\I}{2\bar k}\Delta_T\right\}\varphi(\vec x_T,\zeta-\tau)\ ,
\end{align}
\begin{align}\label{B.35}
\mathcal A^{(1)}(\vec x_T,\zeta,\tau)&=(K\mathcal A^{(0)})(\vec x_T,\zeta,\tau)\nn\\
&=\int^\zeta_{\zeta-\tau}\mathrm d\zeta'\left.(L\mathcal A^{(0)})(\vec x_T,\zeta',\tau')\right|_{\tau'=\tau-\zeta+\zeta'}\nn\\
&=\left\{\kappa_0(\zeta,\tau)
+\kappa_1(\zeta,\tau)\frac{\partial}{\partial\zeta}
+\kappa_2(\zeta,\tau)\frac{\partial^2}{\partial\zeta^2}\right.\nn\\
&\qquad\left.+\frac{\I\tau}{2\bar k}\Delta_T\right\}\mathcal A^{(0)}(\vec x_T,\zeta,\tau)\ .
\end{align}
Here we define
\begin{align}\label{B.36}
\kappa_0(\zeta,\tau)&=\int^\zeta_{\zeta-\tau}\mathrm d\zeta'\,l_0(\mathcal Z(\zeta'),\mathcal Z(\zeta-\tau))\ ,\nn\\
\kappa_1(\zeta,\tau)&=\int^\zeta_{\zeta-\tau}\mathrm d\zeta'\,l_1(\mathcal Z(\zeta'),\mathcal Z(\zeta-\tau))\ ,\nn\\
\kappa_2(\zeta,\tau)&=\int^\zeta_{\zeta-\tau}\mathrm d\zeta'\,l_2(\mathcal Z(\zeta'))\ .
\end{align}
Now we shall make rough estimates of the size of the functions $|\kappa_j(\zeta,\tau)|$, $(j=0,1,2)$. For the parameter $\tau$ to consider we suppose as in (\ref{A.102a})
\begin{align}\label{B.37}
\tau\approx l= 5\,\mathrm m\ .
\end{align}
For the potential we take as an estimate
\begin{align}\label{B.38}
|V(z)|&\lesssim V_m\ ,\nn\\
V_m &= er_B(1){\cal E}_m+\mu_B{\cal B}_m\ ,\nn\\
{\cal E}_m&=10 \,\mathrm{V/cm}\ ,\nn\\
{\cal B}_m&=1\,\mathrm{mT}\ .
\end{align}
With 
\begin{align}\label{B.38a}
\bar k=q_z=5.6\times 10^{10}\,\mathrm m^{-1}\ ,
\end{align}
see (\ref{2.16}), this gives
\begin{align}\label{B.39}
|V(z)|&\lesssim V_m\approx 10^{-7}\,\mathrm{eV},\nn\\
|V(z)|\left(\frac{\bar k^2}{2m}\right)^{-1}&\lesssim V_m
\left(\frac{\bar k^2}{2m}\right)^{-1}\nn\\
&=:\eta=1.5\times 10^{-6}\ .
\end{align}
For the variation of the potential we estimate
\begin{align}\label{B.40}
\left|\frac{\partial V(z)}{\partial z}\right|&\lesssim er_B(1)
\frac{\Delta{\cal E}}{\Delta z}+\mu_B\frac{\Delta{\cal B}}{\Delta z}\ ,\nn\\
\Delta{\cal E}&=1\,\mathrm{V/cm}\ ,\nn\\
\Delta{\cal B}&=0.1\,\mathrm{mT}\ ,\nn\\
\Delta z&=100\,\mathrm{\mu m}\ .
\end{align}
This gives
\begin{align}\label{B.41}
\left|\frac{\partial V(z)}{\partial z}\right|\lesssim \frac{V_m}{10\Delta z}\ .
\end{align}
Similarly we estimate
\begin{align}\label{B.42}
\left|\frac{\partial^2 V(z)}{\partial z^2}\right|\lesssim\frac{V_m}{10\Delta z^2}\ .
\end{align}
For the decay rate we estimate according to (\ref{2.19}) and (\ref{B.37})
\begin{align}\label{B.43}
\Gamma(z)&\lesssim\frac{v_z}{l}=\frac{\bar k}{ml}\ ,\nn\\
\left|\frac{\partial\Gamma(z)}{\partial z}\right|&\lesssim\frac{\bar k}{ml\Delta z}\ .
\end{align}
Now we can estimate $\kappa_j(\zeta,\tau)$, $(j=0,1,2)$, from (\ref{B.36}) and (\ref{B.17})--(\ref{B.19}) as follows
\begin{align}
&|\kappa_0(\zeta,\tau)|\lesssim\frac{\tau}{2\bar k}
\left\{\frac{m^2}{4\bar k^2}\left[\frac{2V_m}{10\bar k\Delta z}+\frac{2\bar k}{ml}\right]^2
\right.\nn\\
&\quad+\frac{m^2}{2\bar k^3}\frac{V_m}{10\Delta z}
\left[\frac{2V_m}{10\bar k\Delta z}+\frac{2\bar k}{ml}\right]\nn\\
&\quad\left.+\frac{m}{\bar k^2}\left[\frac{m}{\bar k^2}\left(\frac{V_m}{10\Delta z}\right)^2
+\frac{V_m}{10\Delta z^2}+\frac{\bar k^2}{ml\Delta z}\right]\right\}\nn\\
&=\frac{l}{2\bar k}\left\{\frac34\left(\frac{\eta}{10\Delta z}\right)^2
+\frac32\frac{\eta}{10 l\Delta z}
+\frac{\eta}{20\Delta z^2}+\frac{1}{l^2}+\frac{1}{l\Delta z}\right\}\nn\\
&\approx\frac{1}{2\bar k\Delta z}\approx 10^{-7}\label{B.44}\ ,\\
&|\kappa_1(\zeta,\tau)|\lesssim \frac{l}{\bar k}\left\{
\frac{m}{\bar k}\left[\frac{1}{\bar k}\frac{V_m}{10\Delta z}+\frac{\bar k}{ml}\right]
+\frac12\frac{m}{\bar k^2}\frac{V_m}{10\Delta z}\right\}\nn\\
&\qquad\qquad=\frac{l}{\bar k}\left\{\frac{1}{l}+\frac{3}{4}\frac{\eta}{10\Delta z}\right\}
\approx\frac{1}{\bar k}\ ,\label{B.45}\\
&|\kappa_2(\zeta,\tau)|\lesssim\frac{l}{2\bar k}\ .\label{B.46}
\end{align}
With (\ref{B.44})--(\ref{B.46}) we can estimate the size of $\mathcal A^{(1)}$ relative to $A^{(0)}$ from (\ref{B.35}). We choose as a concrete example for the amplitude $\mathcal A^{(0)}$ the one corresponding to the Gaussian function $\chi$ in (\ref{3.29}). From (\ref{3.27}), (\ref{3.29}) and (\ref{B.33}) we get
\begin{align}\label{B.47}
&{\mathcal A}^{(0)}(\bx_T,\zeta,\tau)=\varphi(\bx_T,\zeta-\tau)=
{\cal N}(2\pi)^{-3/4}\sigma_T^{-1}\sigma^{-1/2}_L\nn\\
&\qquad\times\exp\left[-\frac{1}{4\sigma^2_T}\bx^2_T-\frac{1}{4\sigma^2_L}(\zeta-\tau)^2\right]\nn\\
&\qquad\times\exp\left.\left[\I\int^u_{z_0}dz'\frac{2mV(z')}{k(z')+\bar k}-\I\gamma(u)\right]\right|_{u=\mathcal Z(\zeta-\tau)}\ .
\end{align}
Here we suppose
\begin{align}\label{B.48}
100\,\mu \mathrm m<\sigma_L,\sigma_T<300\,\mu \mathrm m\ .
\end{align}
The Gaussian function in (\ref{B.47}) is nonzero practically only for 
\begin{align}\label{B.49}
|\zeta-\tau|\lesssim 3\sigma_L\ .
\end{align}
Then we get also for $u=\mathcal Z(\zeta-\tau)$ from (\ref{3.3}) and (\ref{3.3a})
\begin{align}\label{B.50}
|u-z_0|\lesssim 3\sigma_L\ .
\end{align}
This means that $u$ in (\ref{B.47}) varies only in the field-free region around $z_0$. Thus, the last exponential function on the r.h.s. of (\ref{B.47}) can be replaced by 1 and, similarly, we get $\mathcal N=1$ from (\ref{3.30}). The amplitude $\mathcal A^{(0)}$ and its derivatives with respect to $\zeta$ read then
\begin{align}\label{B.51}
{\mathcal A}^{(0)}(\bx_T,\zeta,\tau)
&=(2\pi)^{-3/4}\sigma^{-1}_T\sigma^{-1/2}_L\nn\\
&\quad\times\exp\left[-\frac{1}{4\sigma^2_T}\bx^2_T-\frac{1}{4\sigma^2_L}(\zeta-\tau)^2\right]\ ,
\end{align}
\begin{align}\label{B.52}
\frac{\partial {\mathcal A}^{(0)}}{\partial\zeta}(\bx_T,\zeta,\tau)
=-\frac{1}{2\sigma^2_L}(\zeta-\tau){\mathcal A}^{(0)}(\bx_T,\zeta,\tau)\ ,
\end{align}
\begin{align}\label{B.53}
\frac{\partial^2{\mathcal A}^{(0)}}{\partial\zeta^2}(\bx_T,\zeta,\tau)=
-\frac{1}{2\sigma^2_L}\left[1-\frac{(\zeta-\tau)^2}{2\sigma^2_L}\right]
{\mathcal A}^{(0)}(\bx_T,\zeta,\tau)\ ,
\end{align}
\begin{align}\label{B.54}
\Delta_T{\mathcal A}^{(0)}(\bx_T,\zeta,\tau)=
-\frac{1}{2\sigma^2_T}\left(2-\frac{\bx^2_T}{2\sigma^2_T}\right)
{\mathcal A}^{(0)}(\bx_T,\zeta,\tau)\ .
\end{align}
Inserting (\ref{B.51})--(\ref{B.54}) in (\ref{B.35}) we get
\begin{align}\label{B.55}
{\mathcal A}^{(1)}(\bx_T,\zeta,\tau)=&
\left\{\kappa_0(\zeta,\tau)
-\kappa_1(\zeta,\tau)\frac{1}{2\sigma^2_L}(\zeta-\tau)\right.\nn\\
&-\kappa_2(\zeta,\tau)
\frac{1}{2\sigma^2_L}\left[1-\frac{(\zeta-\tau)^2}{2\sigma^2_L}\right]\nn\\
&\left.-\frac{\I\tau}{2\bar k}\frac{1}{2\sigma^2_T}\left(2-\frac{\bx^2_T}{2\sigma_T^2}\right)\right\}
{\mathcal A}^{(0)}(\bx_T,\zeta,\tau)\ .
\end{align}
In the only interesting range given by (\ref{B.49}) we get with (\ref{B.44})--(\ref{B.46}) as an estimate of the order of magnitudes
\begin{align}\label{B.55a}
\left|\frac{\mathcal A^{(1)}(\vec x_T,\zeta,\tau)}{\mathcal A^{(0)}(\vec x_T,\zeta,\tau)}\right|
&\lesssim\left\{|\kappa_0(\zeta,\tau)|+|\kappa_1(\zeta,\tau)|\frac{1}{2\sigma_L}\right.\nn\\
&\quad\left.+|\kappa_2(\zeta,\tau)|\frac{1}{2\sigma^2_L}
+\frac{l}{2\bar k}\frac{1}{\sigma^2_T}\right\}\nn\\
&\approx 10^{-7}+\frac{3}{2\bar k\sigma_L}
+\frac{l}{4\bar k\sigma^2_L}+\frac{l}{2\bar k\sigma^2_T}\ .
\end{align}
where the quantity $\vec x_T^2/(2\sigma_T^2)$ from the last term on the r.h.s. of (\ref{B.55}) is neglected due to the exponential decrease of $\mathcal A^{(0)}(\vec x_T,\zeta,\tau)$ for increasing $|\vec x_T|$.
Numerically the last two terms on the r.h.s. of (\ref{B.55a}) are the dominant ones and we find
\begin{align}\label{B.56}
\left|\frac{{\mathcal A}^{(1)}(\bx_T,\zeta,\tau)}{{\mathcal A}^{(0)}(\bx_T,\zeta,\tau)}\right|\lesssim
\frac{l}{2\bar k}\left(\frac{1}{2\sigma^2_L}+\frac{1}{\sigma^2_T}\right)\ .
\end{align}
Inserting here $l,\bar k$ and $\sigma_L,\sigma_T$ from (\ref{2.15}), (\ref{B.38a}) and (\ref{B.48}), respectively, we get
\begin{align}\label{B.57}
\frac{l}{2\bar k}\left(\frac{1}{2\sigma^2_L}+\frac{1}{\sigma^2_T}\right)\lesssim 0.01\ .
\end{align}

To summarize: we find that with our conditions specified above the zero order amplitude ${\mathcal A}^{(0)}$ is accurate with errors of at most $1\%$. The origin of the dominant errors shown in (\ref{B.56}) is easily identified. It is due to the fact that ${\mathcal A}^{(0)}$ does not include the longitudinal and transverse dispersion of the wave packet. But this presents no problem, since we have discussed in Section \ref{s:SolutionOneComp} and in this appendix only wavelets, where we can - within reasonable limits - choose the widths $\sigma_L$ and $\sigma_T$ as large as we want. The actual narrow wave packets are formed as superpositions of the wavelets as discussed in Section \ref{s:InterferenceEffects} and Appendix \ref{s:AppendixC}. Finally we emphasise again that we have given here only an \textit{example} of accuracy estimates for the wavelet solutions of (\ref{B.15}). For other experimental conditions, that is other electric and magnetic fields, other atom velocities etc., similar estimates can easily be performed.

%========================================================================================
\section{Interference effects, details}\label{s:AppendixC}
%========================================================================================
 \setcounter{equation}{0}

Here we represent the details of the calculations of the interference effects discussed in Section \ref{s:InterferenceEffects}. The $z$ component of the probability current for the wave function $\ket{\psi_p(\bx,t)}$ (\ref{4.18}) is given in (\ref{4.22}). Inserting there (\ref{4.20}) and taking into account that the $\bar k$ distribution is supposed to be sharply peaked at $\bar k_m$, see (\ref{4.5a}), leads to (\ref{4.23}). This gives for the integrated flux ${\cal F}_p$ (\ref{4.21}) the result
\begin{align}\label{C.1}
{\cal F}_p&=\frac{\bar k_m}{m}\int \mathrm dt\int \mathrm d^2x_T\int\frac{\mathrm d\bar k'\mathrm d\bar k}{(2\pi)^2}
f^*(\bar k')f(\bar k)\nn\\
&\quad\times\exp\left[-\I\frac{\bar k^2-\bar k'^2}{2m}t+i(\bar k-\bar k')(z_a-z_0)\right]\nn\\
&\quad\times\sum_{\alpha,\beta\in I}p_\beta p^*_\alpha c^*_\beta c_\alpha
U^*_\beta(z_a,u;\bar k')|_{u=\mathcal Z_\beta(\zeta_\beta-\frac{\bar k'}{m}t)}\nn\\
&\qquad\qquad \times U_\alpha(z_a,u;\bar k)|_{u=\mathcal Z_\alpha(\zeta_\alpha-\frac{\bar k}{m}t)}\nn\\
&\quad\times\chi^*(\bx_T,\zeta_\beta-\frac{\bar k'}{m}t)\,\chi(\bx_T,\zeta_\alpha-\frac{\bar k}{m}t)
\end{align}
where
\begin{align}\label{C.2}
\zeta_\alpha=\frak Z_\alpha(z_a)~,~\zeta_\beta={\frak Z}_\beta(z_a)\ .
\end{align}
Now we make the following approximations
\begin{align}\label{C.3}
\frac{\bar k^2-\bar k'^2}{2m}t=
(\bar k-\bar k')\frac{\bar k+\bar k'}{2m}t
&\approx (\bar k-\bar k')\frac{\bar k_m}{m}t\ ,\nn\\
\exp\left[-\I\frac{\bar k^2-\bar k'^2}{2m}t+\I(\bar k-\bar k')(z_a-z_0)\right]
&\approx\exp\left[ i(\bar k-\bar k')z'\right]
\end{align}
with
\begin{align}\label{C.4}
z'=z_a-z_0-\frac{\bar k_m}{m}t\ .
\end{align}
Furthermore we have
\begin{align}\label{C.5}
\zeta_\alpha-\frac{\bar k}{m}t={\frak Z}_\alpha(z_a)-\frac{\bar k}{m}t
=\Delta\tau_\alpha(z_a;\bar k)+z_a-z_0-\frac{\bar k}{m}t\ .
\end{align}
Here $\Delta\tau_\alpha(z_a;\bar k)$ gives according to (\ref{3.18}) the difference of reduced arrival times of the wave (\ref{4.14}) relative to the free wave. We have with (\ref{3.3}), (\ref{3.3a}) and (\ref{3.21})
\begin{align}\label{C.5a}
\Delta\tau_\alpha(z_a;\bar k)&=\int^{z_a}_{z_0}\mathrm dz'
\left(\frac{\bar k}{k_\alpha(z')}-1\right)\nn\\
&=\int^{z_a}_{z_0}\mathrm dz'\frac{2mV_\alpha(z')}{k_\alpha(z')(k_\alpha(z')+\bar k)}\ .
\end{align}
The function
\begin{align}\label{C.5b}
\chi\left(\bx_T,\zeta_\alpha-\frac{\bar k}{m}t\right)
\end{align}
is significantly non zero only for
\begin{align}\label{C.7}
\left|\zeta_\alpha-\frac{\bar k}{m}t\right|\lesssim 3\sigma_L
\end{align}
with $\sigma_L=100$ to $300\,\mu \mathrm m$. Correspondingly,
\begin{align}\label{C.8}
u=\mathcal Z_\alpha\left(\zeta_\alpha-\frac{\bar k}{m}t\right)
\end{align}
varies only over a distance of a few $\sigma_L$ around $z_0$, that is, in the field-free region
\begin{align}\label{C.8a}
|u-z_0|\lesssim 3\sigma_L\ .
\end{align}
We consider now (see (\ref{3.22}))
\begin{align}\label{C.9}
U_\alpha(z_a,u;\bar k)
=\exp\left[-\I\int^{z_a}_u\mathrm dz'\frac{2mV_\alpha(z')}{k_\alpha(z')+\bar k}
-\frac12\ln \frac{k_\alpha(z_a)}{k_\alpha(u)}\nn\right.\\
\left.-\frac12 \int^{z_a}_u\mathrm dz'\frac{m\Gamma_\alpha(z')}{k_\alpha(z')}
+\I\gamma_{\alpha\alpha}(z_a)-\I\gamma_{\alpha\alpha}(u)\right]
\end{align}
for $u$ given by (\ref{C.8}), (\ref{C.8a}). We have then
\begin{align}\label{C.9a}
k_\alpha(z_a)=k_\alpha(u)=\bar k\, ,\qquad \ln\frac{k_\alpha(z_a)}{k_\alpha(u)}=0\ ,
\end{align}
\begin{align}\label{C.9b}
\int^{z_a}_u\mathrm dz'\frac{2mV_\alpha(z')}{k_\alpha(z')+\bar k}
=\int^{z_a}_{z_0}\mathrm dz'\frac{2mV_\alpha(z')}{k_\alpha(z')+\bar k}\ ,
\end{align}
\begin{align}\label{C.9c}
\gamma_{\alpha\alpha}(u)=\gamma_{\alpha\alpha}(z_0)\ .
\end{align}
Also the error made by setting
\begin{align}\label{C.10}
\int^{z_a}_u\mathrm dz'\frac{m\Gamma_\alpha(z')}{k_\alpha(z')}\approx
\int^{z_a}_{z_0}\mathrm dz'\frac{m\Gamma_\alpha(z')}{k_\alpha(z')}
\end{align}
is negligible. Therefore we can set
\begin{align}
&U_\alpha(z_a,u;\bar k)|_{u=\mathcal Z_\alpha\left(\zeta_\alpha-\frac{\bar k}{m}t\right)}
=U_\alpha(z_a,z_0;\bar k)\nn\\
&\quad=\exp\left[-\I\int^{z_a}_{z_0}\mathrm dz'\frac{2mV_\alpha(z')}{k_\alpha(z')+\bar k}\right.\nn\\
&\quad\left.-\frac12\int^{z_a}_{z_0}\mathrm dz'\frac{m\Gamma_\alpha(z')}{k_\alpha(z')}
+\I\gamma_{\alpha\alpha}(z_a)-\I\gamma_{\alpha\alpha}(z_0)\right]\ .\label{C.11}
\end{align}
Now we remember that $f(\bar k)$ in the integrand in (\ref{C.1}) is sharply peaked for $\bar k=\bar k_m$. We expand therefore the terms in the argument of the exponential function in (\ref{C.11}) around $\bar k_m$. For the integral over $V_\alpha(z')$ which gives a fast oscillating term we expand up to first order in $(\bar k-\bar k_m)$. For the remaining terms we keep only the zero-order contributions. We get then
\begin{align}\label{C.15}
&\int^{z_a}_{z_0}\mathrm dz'\frac{2mV_\alpha(z')}{k_\alpha(z')+\bar k}=
\int^{z_a}_{z_0}\mathrm dz'\left.\frac{2mV_\alpha(z')}{k_\alpha(z')+\bar k}\right|_{\bar k=\bar k_m}\nn\\
&\qquad-(\bar k-\bar k_m)\int^{z_a}_{z_0}\mathrm dz'\left.\frac{2mV_\alpha(z')}{k_\alpha(z')\big(k_\alpha(z')+\bar k\big)}\right|_{\bar k=\bar k_m}\nn\\
&\qquad+\mathcal O\big[(\bar k-\bar k_m)^2\big]\nn\\
&\quad=\int^{z_a}_{z_0}\mathrm dz'\left.\frac{2mV_\alpha(z')}{k_\alpha(z')+\bar k}\right|_{\bar k=\bar k_m}-(\bar k-\bar k_m)\Delta\tau_\alpha\nn\\
&\qquad+\mathcal O\big[(\bar k-\bar k_m)^2\big]
\end{align}
where we set
\begin{align}\label{C.16}
\Delta\tau_\alpha\equiv\Delta\tau_\alpha(z_a;\bar k_m)\ ,
\end{align}
see (\ref{C.5a}) and (\ref{4.25}). Inserting (\ref{C.15}) in (\ref{C.11}) we find with the approximations discussed above. 
\begin{align}\label{C.17}
U_\alpha(z_a,z_0;\bar k)\approx
U_\alpha(z_a,z_0;\bar k_m)
\exp\big[\I(\bar k-\bar k_m)\Delta\tau_\alpha\big]\ ,
\end{align}
\begin{align}\label{C.18}
&U_\alpha(z_a,z_0;\bar k_m)\approx
\exp\left[-\I\int^{z_a}_{z_0}\mathrm dz'\frac{mV_\alpha(z')}{\bar k_m}\right.\nn\\
&\qquad\left.\quad-\frac12\int^{z_a}_{z_0}\mathrm dz'\frac{m\Gamma_\alpha(z')}{\bar k_m}
+\I\gamma_{\alpha\alpha}(z_a)-\I\gamma_{\alpha\alpha}(z_0)\right]\nn\\
&\quad=\exp[-\I\phi_{\mathrm{dyn},\alpha}+\I\phi_{\mathrm{geom},\alpha}]\ .
\end{align}
Here we use the definitions (\ref{4.28}) and (\ref{4.29}). Putting now everything together we find from (\ref{C.1})--(\ref{C.5}) and (\ref{C.17}), (\ref{C.18})
\begin{align}\label{C.19}
&{\cal F}_p=\int\frac{\mathrm d\bar k'\mathrm d\bar k}{(2\pi)^2}f^*(\bar k')f(\bar k)
\int \mathrm dz'\int \mathrm d^2x_T\exp\big[\I(\bar k-\bar k')z'\big]\nn\\
&\times\sum_{\alpha,\beta\in I}p_\beta p_\alpha^*c^*_\beta c_\alpha\exp\big[-\I(\bar k'-\bar k_m)\Delta\tau_\beta+\I(\bar k-\bar k_m)\Delta\tau_\alpha\big]\nn\\
&\times\chi^*(\bx_T,z'+\Delta\tau_\beta)\,\chi(\bx_T,z'+\Delta\tau_\alpha)\nn\\
&\times U_\beta^*(z_a,z_0;\bar k_m)U_\alpha(z_a,z_0;\bar k_m)\ .
\end{align}
Here and in the following all integrals run from $-\infty$ to $\infty$. Now we change variables and set
\begin{align}\label{C.20}
\bar k_s&=\frac12\bar k+\frac12\bar k'\ , \nn\\
\bar k_d&=\bar k-\bar k'\ .
\end{align}
This gives (\ref{4.24}) with $g(\bar k_s,\Delta\tau_\beta,\Delta\tau_\alpha)$ as defined in (\ref{4.26}). Inserting there the Fourier transform of $\chi$ according to (\ref{4.5}) we get
\begin{align}\label{C.21}
&g(\bar k_s,\Delta\tau_\beta,\Delta\tau_\alpha)\nn\\
&\quad=\frac{1}{(2\pi)^4}\int \mathrm d^2q_T\,\mathrm dq'_L\,\mathrm dq_L\,f^*\big(\bar k_s-\frac12(q'_L-q_L)\big)\nn\\
&\qquad\times f\big(\bar k_s+\frac12(q'_L-q_L)\big)
\exp\big[\I(q'_L+q_L)\frac12(\Delta\tau_\alpha-\Delta\tau_\beta)\big]\nn\\
&\qquad\times\tilde \chi^*(\vec q_T,q'_L)\,\tilde \chi(\vec q_T,q_L)\ .
\end{align}
Finally, we insert the explicit Gaussian functions $f(\cdot)$ and $\tilde \chi(\cdot)$ from (\ref{4.9}) and (\ref{4.10}), respectively, in (\ref{C.21}) and get for this case
\begin{align}\label{C.22}
&g(\bar k_s,\Delta\tau_\beta,\Delta\tau_\alpha)=
\sqrt{2\pi}2\sigma_k\nn\\
&\quad\times\exp\left[-2\sigma^2_k(\bar k_s-\bar k_m)^2
-\frac{1}{8\sigma^2_L}(\Delta\tau_\alpha-\Delta\tau_\beta)^2\right]\ .
\end{align}
With (\ref{C.22}) we get for the integrated flux ${\cal F}_p$ from (\ref{4.24}) the final result (\ref{4.31}).

\onecolumn

%========================================================================================
\section{Matrices for $n=2$ states of hydrogen}\label{s:AppendixD}
%========================================================================================
 \setcounter{equation}{0}

Tables \ref{t:H2.M0}, \ref{t:H2.D} and \ref{t:H2.Mu} show the non-zero parts of the mass matrix $\umat{\tilde M}_0(\delta_1,\delta_2)$ for zero external fields, of the electric dipole operator $\uvec{D}$ and of the magnetic dipole operator $\uvec{\mu}$ for the $n=2$ states of hydrogen. We give all these matrices in the basis of the pure 2S and 2P states, that is the states for zero external fields and without the P-violating mixing. Thus, the P-violation parameters $\delta_{1,2}$ occur in the matrix $\umat{\tilde M}(\delta_1,\delta_2)$ explicitly.

In Tables \ref{t:H2.D} and \ref{t:H2.Mu} we use the spherical unit vectors, which are defined as
\begin{align}
\vec e_0 = \vec e_3\ ,\qquad\vec e_\pm = \mp\frac1{\sqrt2}\klr{\vec e_1 \pm \I\vec e_2}\ ,
\end{align}
where $\vec e_i$ $(i=1,2,3)$ are the Cartesian unit vectors. For $\vec e_\pm$, the following relation holds:
\begin{align}
\vec e_\pm^* = -\vec e_\mp\ .
\end{align}

\begin{table}[htbp]
\begin{center}
\caption{The mass matrix $\umat{\tilde M}_0(\delta_1,\delta_2)$ (\ref{2.4b}) for the $n=2$ states of hydrogen for the case of zero external fields. For the explanation of the variables $\Delta$, $L$ and $\mcal A$ and their numerical values see the introduction of Appendix \ref{s:AppendixA} and Table \ref{t:values}. The PV parameters $\delta_{1,2}$ can also be found in Table \ref{t:values}, the decay rates $\Gamma_{P,S}$ are given in (\ref{2.18}).
% Here, $\Delta$ fine structure splitting energy, $\mcal A$ is the hyperfine splitting constant and $\LambShift$ is the $E(2S_{1/2})-E(2P_{1/2})$ Lamb shift of hydrogen.
}\label{t:H2.M0}
\vskip10pt
\begin{tabular}{l||c|cccc|}
& \parbox{18mm}{\vskip5pt$\ 2P_{3/2},2,2\ $\vskip4pt} & $\ 2P_{3/2},2,1\ $ & $\ 2P_{3/2},1,1\ $ & $\ 2P_{1/2},1,1\ $ & $\ 2S_{1/2},1,1\ $\\  \hline\hline
%& $2,2$ & $2,1$ & $1,1$ & $1,1$ & $1,1$\\  \hline\hline
$\ 2P_{3/2},2,2\ $ & \mycell{18mm}{$\FineStructure+\frac{\HyperFineSplitting}{160}$\\[-0.0ex]$-\tfrac{\I }{2}\Gamma_P$}
& 0 & 0 & 0 & 0\\ \hline
$\ 2P_{3/2},2,1\ $ & 0 & 
\mycell{18mm}{${\FineStructure}+\frac{\HyperFineSplitting}{160}$\\[-0.0ex]$-\frac{\I}{2}\,{\Gamma_P}$}
& 0 & 0 & 0\\  
$\ 2P_{3/2},1,1\ $ & 0 & 0 & 
\mycell{18mm}{${\FineStructure}-\frac{\HyperFineSplitting}{96}$\\[-0.0ex]$-\frac{\I}{2}\,{\Gamma_P}$} & 
$-\frac{\HyperFineSplitting}{192\,{\sqrt{2}}}$ & 0\\  
$\ 2P_{1/2},1,1\ $ & 0 & 0 & 
$-\frac{\HyperFineSplitting}{192\,{\sqrt{2}}}$ & 
$\frac{\HyperFineSplitting}{96}-\frac\I2{\Gamma_P}$ & 
\mycell{18mm}{$\I{\delta_1}{\LambShift}$\\[-0.0ex]$+\frac\I2\delta_2 \LambShift$} \\ 
$\ 2S_{1/2},1,1\ $ & 0 & 0 & 0 & 
\mycell{18mm}{$-\I{\delta_1}{\LambShift}$\\[-0.0ex]$-\frac\I2 {\delta_2}{\LambShift}$} &
\mycell{18mm}{${\LambShift}+\frac{\HyperFineSplitting}{32}$\\[-0.0ex]$-\frac{\I}{2}\,{\Gamma_S}$}\\  \hline
\end{tabular}\\[15pt]
\centerline{\bf (Table \ref{t:H2.M0}a)}
\end{center}
\end{table}
\begin{table}
\begin{center}
\begin{tabular}{l||cccccc|}
& \parbox{18mm}{\vskip5pt$\ 2P_{3/2},2,0\ $\vskip4pt} & $\ 2P_{3/2},1,0\ $ & $\ 2P_{1/2},1,0\ $ & $\ 2S_{1/2},1,0\ $ & $\ 2P_{1/2},0,0\ $ & $\ 2S_{1/2},0,0\ $\\ \hline\hline
%& $2,0$ & $1,0$ & $1,0$ & $1,0$ & $0,0$ & $0,0$\\ \hline
$\ 2P_{3/2},2,0\ $ & 
\mycell{18mm}{${\FineStructure}+\frac{\HyperFineSplitting}{160}$\\[-0.0ex]$-\frac{\I}{2}{\Gamma_P}$} & 0 & 0 & 0
& 0 & 0\\  
$\ 2P_{3/2},1,0\ $ & 0 & 
\mycell{18mm}{${\FineStructure}-\frac{\HyperFineSplitting}{96}$\\[-0.0ex]$-\frac{\I}{2}{\Gamma_P}$} & $-\frac{\HyperFineSplitting}{192{\sqrt{2}}}$ & 0 & 0 & 0 \\  
$\ 2P_{1/2},1,0\ $ & 0 & 
$-\frac{\HyperFineSplitting}{192{\sqrt{2}}}$ & $\frac{\HyperFineSplitting}{96} - \frac\I2{\Gamma_P}$ & 
\mycell{18mm}{$\I\delta_1{\LambShift}$\\[-0.0ex]$+\frac\I2\delta_2{\LambShift}$} & 0 & 0 \\ 
$\ 2S_{1/2},1,0\ $ & 0 & 0 & 
\mycell{18mm}{$-\I{\delta_1}{\LambShift}$\\[-0.0ex]$-\frac\I2{\delta_2}{\LambShift}$} & 
\mycell{18mm}{${\LambShift}+\frac{\HyperFineSplitting}{32}$\\[-0.0ex]$-\frac{\I}{2}{\Gamma_S}$} & 0 & 0 \\ 
$\ 2P_{1/2},0,0\ $ & 0 & 0 & 0 & 0 & 
$-\frac{\HyperFineSplitting}{32} - \frac\I2{\Gamma_P}$ & 
\mycell{18mm}{$\I{\delta_1}{\LambShift}$\\[-0.0ex]$+\frac{3}{2}\I{\delta_2}{\LambShift}$} \\
$\ 2S_{1/2},0,0\ $ & 0 & 0 & 0 & 0 & 
\mycell{18mm}{$-\I{\delta_1}{\LambShift}$\\[-0.0ex]$-\frac{3}{2}\I{\delta_2}{\LambShift}$} & 
\mycell{18mm}{${\LambShift}-\frac{3\HyperFineSplitting}{32}$\\[-0.0ex]$-\frac{\I}{2}{\Gamma_S}$} \\ \hline
\end{tabular}\\[15pt]
\centerline{\bf (Table \ref{t:H2.M0}b)}
\end{center}
\end{table}
\begin{table}
\begin{center}
\begin{tabular}{l||cccc|c|}
& \parbox{20mm}{\vskip5pt$\ 2P_{3/2},2,-1\ $\vskip4pt} & $\ 2P_{3/2},1,-1\ $ & $\ 2P_{1/2},1,-1\ $ & $\ 2S_{1/2},1,-1\ $ & $\ 2P_{3/2},2,-2\ $\\ \hline\hline
%& $2,-1$ & $1,-1$ & $1,-1$ & $1,-1$ & $2,-2$\\ \hline\hline
$\ 2P_{3/2},2,-1\ $ & 
\mycell{18mm}{${\FineStructure}+\frac{\HyperFineSplitting}{160}$\\[-0.0ex]$-\frac{\I}{2}{\Gamma_P}$} 
& 0 & 0 & 0 & 0 \\  
$\ 2P_{3/2},1,-1\ $ & 0 & 
\mycell{18mm}{${\FineStructure}-\frac{\HyperFineSplitting}{96}$\\[-0.0ex]$-\frac{\I}{2}{\Gamma_P}$} & 
$-\frac{\HyperFineSplitting}{192{\sqrt{2}}}$ & 0 & 0 \\ 
$\ 2P_{1/2},1,-1\ $ & 0 & 
$-\frac{\HyperFineSplitting}{192{\sqrt{2}}}$ & 
$\frac{\HyperFineSplitting}{96} - \frac\I2{\Gamma_P}$ & 
\mycell{18mm}{$\I{\delta_1}{\LambShift}$\\[-0.0ex]$+\frac\I2{\delta_2}{\LambShift}$} & 0 \\ 
$\ 2S_{1/2},1,-1\ $ & 0 & 0 & 
\mycell{18mm}{$-\I{\delta_1}{\LambShift}$\\[-0.0ex]$-\frac\I2{\delta_2}{\LambShift}$} & 
\mycell{18mm}{${\LambShift}+\frac{\HyperFineSplitting}{32}$\\[-0.0ex]$-\frac{\I}{2}{\Gamma_S}$} & 0 \\  \hline
$\ 2P_{3/2},2,-2\ $ & 0 & 0 & 0 & 0 & 
\mycell{18mm}{${\FineStructure}+\frac{\HyperFineSplitting}{160}$\\[-0.0ex]$-\frac{\I}{2}{\Gamma_P}$} \\ \hline
\end{tabular}\\[15pt]
\centerline{\bf (Table \ref{t:H2.M0}c)}
\end{center}
\end{table}

%\begin{turnpage}
{
\begin{table}
\begin{center}
\caption{The suitably normalised electric dipole operator $\uvec{D}/(e\,r_B(1))$ for the $n=2$ states of hydrogen.}\label{t:H2.D}
\vskip10pt
\begin{tabular}{l||c|cccc|}
& \mycell{18mm}{$\ 2P_{3/2},2,2\ $} & $\ 2P_{3/2},2,1\ $ & $\ 2P_{3/2},1,1\ $ & $\ 2P_{1/2},1,1\ $ & $\ 2S_{1/2},1,1\ $\\ 
%& $2,2$ & $2,1$ & $1,1$ & $1,1$ & $1,1$\\ 
\hline\hline
\mycell{18mm}{$\ 2P_{3/2},2,2\ $} & 0 & 0 & 0 & 0 & $-3\sem$\\ \hline
\mycell{18mm}{$\ 2P_{3/2},2,1\ $} & 0 & 0 & 0 & 0 & 
$\frac{3}{\sqrt{2}}\sez$\\
\mycell{18mm}{$\ 2P_{3/2},1,1\ $} & 0 & 0 & 0 & 0 & 
$-{\sqrt{\frac{3}{2}}}\sez$\\
\mycell{18mm}{$\ 2P_{1/2},1,1\ $} & 0 & 0 & 0 & 0 & 
$-{\sqrt{3}}\sez$\\
\mycell{18mm}{$\ 2S_{1/2},1,1\ $} & $3\sep$ & 
$\frac{3}{{\sqrt{2}}}\sez$ & 
$-{\sqrt{\frac{3}{2}}}\sez$ & 
$-{\sqrt{3}}\sez$ & 0\\ \hline
\end{tabular}\\[15pt]
\centerline{\bf (Table \ref{t:H2.D}a)}
\vskip15pt
% \end{table}
% \begin{table}
\begin{tabular}{l||cccccc|}
& \mycell{18mm}{$\ 2P_{3/2},2,0\ $} & $\ 2P_{3/2},1,0\ $ & $\ 2P_{1/2},1,0\ $ & $\ 2S_{1/2},1,0\ $ & $\ 2P_{1/2},0,0\ $ & $\ 2S_{1/2},0,0\ $\\ \hline\hline
\mycell{18mm}{$\ 2P_{3/2},2,1\ $} & 0 & 0 & 0 & 
$-\frac{3}{\sqrt2}\sem$ & 0 & 0\\ 
\mycell{18mm}{$\ 2P_{3/2},1,1\ $} & 0 & 0 & 0 & 
$-\sqrt{\frac{3}{2}}\sem$ & 0 & 
$-\sqrt{6}\sem$\\ 
\mycell{18mm}{$\ 2P_{1/2},1,1\ $} & 0 & 0 & 0 & 
$-\sqrt{3}\sem$ & 0 & 
$\sqrt{3}\sem$\\ 
\mycell{18mm}{$\ 2S_{1/2},1,1\ $} & 
$\sqrt{\frac{3}{2}}\sem$ & 
$-\sqrt{\frac{3}{2}}\sem$ & 
$-\sqrt{3}\sem$ & 0 & 
$\sqrt{3}\sem$ & 0\\ \hline
\end{tabular}\\[15pt]
\centerline{\bf (Table \ref{t:H2.D}b)}
\end{center}
\end{table}
\begin{table}
\begin{center}
\begin{tabular}{l||cccc}
& \mycell{18mm}{$\ 2P_{3/2},2,1\ $} & $\ 2P_{3/2},1,1\ $ & $\ 2P_{1/2},1,1\ $ & $\ 2S_{1/2},1,1\ $\\ \hline\hline
\mycell{18mm}{$\ 2P_{3/2},2,0\ $} & 0 & 0 & 0 & $-\sqrt{\frac32}\sep$ \\ 
\mycell{18mm}{$\ 2P_{3/2},1,0\ $} & 0 & 0 & 0 & $\sqrt{\frac32}\sep$ \\ 
\mycell{18mm}{$\ 2P_{1/2},1,0\ $} & 0 & 0 & 0 & $\sqrt3\sep$ \\ 
\mycell{18mm}{$\ 2S_{1/2},1,0\ $} & $\frac3{\sqrt2}\sep$ & $\sqrt{\frac32}\sep$ & $\sqrt3\sep$ & 0 \\ 
\mycell{18mm}{$\ 2P_{1/2},0,0\ $} & 0 & 0 & 0 & $-\sqrt3\sep$ \\ 
\mycell{18mm}{$\ 2S_{1/2},0,0\ $} & 0 & $\sqrt6\sep$ & $-\sqrt3\sep$ & 0 \\ \hline
\end{tabular}\\[15pt]
\centerline{\bf (Table \ref{t:H2.D}c)}
\vskip15pt
% \end{table}
% \begin{table}
\begin{tabular}{l||cccccc|}
& \mycell{18mm}{$\ 2P_{3/2},2,0\ $} & $\ 2P_{3/2},1,0\ $ & $\ 2P_{1/2},1,0\ $ & $\ 2S_{1/2},1,0\ $ & $\ 2P_{1/2},0,0\ $ & $\ 2S_{1/2},0,0\ $\\ \hline\hline
%& $2,0$ & $1,0$ & $1,0$ & $1,0$ & $0,0$ & $0,0$\\ \hlx{vhhv}
\mycell{18mm}{$\ 2P_{3/2},2,0\ $} & 0 & 0 & 0 & 
${\sqrt{6}}\sez$ & 0 & 0\\
\mycell{18mm}{$\ 2P_{3/2},1,0\ $} & 0 & 0 & 0 & 0 & 0 & 
${\sqrt{6}}\sez$\\
\mycell{18mm}{$\ 2P_{1/2},1,0\ $} & 0 & 0 & 0 & 0 & 0 & 
$-{\sqrt{3}}\sez$\\
\mycell{18mm}{$\ 2S_{1/2},1,0\ $} & 
${\sqrt{6}}\sez$ & 0 & 0 & 0 & 
$-{\sqrt{3}}\sez$ & 0\\
\mycell{18mm}{$\ 2P_{1/2},0,0\ $} & 0 & 0 & 0 & 
$-{\sqrt{3}}\sez$ & 0 & 0\\
\mycell{18mm}{$\ 2S_{1/2},0,0\ $} & 0 & 
${\sqrt{6}}\sez$ & 
$-{\sqrt{3}}\sez$ & 0 & 0 & 0\\ \hline
\end{tabular}\\[15pt]
\centerline{\bf (Table \ref{t:H2.D}d)}
\end{center}
\end{table}
\begin{table}
\begin{center}
\begin{tabular}{l||cccc|}
& \mycell{18mm}{$\ 2P_{3/2},2,-1\ $} & $\ 2P_{3/2},1,-1\ $ & $\ 2P_{1/2},1,-1\ $ & $\ 2S_{1/2},1,-1\ $\\ \hline\hline
%& $2,-1$ & $1,-1$ & $1,-1$ & $1,-1$\\ \hlx{vhhv}
\mycell{18mm}{$\ 2P_{3/2},2,0\ $} & 0 & 0 & 0 & 
$-\sqrt{\frac{3}{2}}\sem$\\ 
\mycell{18mm}{$\ 2P_{3/2},1,0\ $} & 0 & 0 & 0 & 
$-\sqrt{\frac{3}{2}}\sem$\\
\mycell{18mm}{$\ 2P_{1/2},1,0\ $} & 0 & 0 & 0 & 
$-\sqrt{3}\sem$\\
\mycell{18mm}{$\ 2S_{1/2},1,0\ $} & 
$\frac{3}{\sqrt2}\sem$ & 
$-\sqrt{\frac{3}{2}}\sem$ & 
$-\sqrt{3}\sem$ & 0\\
\mycell{18mm}{$\ 2P_{1/2},0,0\ $} & 0 & 0 & 0 & 
$-\sqrt{3}\sem$\\
\mycell{18mm}{$\ 2S_{1/2},0,0\ $} & 0 & 
$\sqrt{6}\sem$ & 
$-\sqrt{3}\sem$ & 0\\ \hline
\end{tabular}\\[15pt]
\centerline{\bf (Table \ref{t:H2.D}e)}\vskip15pt
% \end{table}
% \begin{table}
\begin{tabular}{l||cccccc|}
& \mycell{18mm}{$\ 2P_{3/2},2,0\ $} & $\ 2P_{3/2},1,0\ $ & $\ 2P_{1/2},1,0\ $ & $\ 2S_{1/2},1,0\ $ & $\ 2P_{1/2},0,0\ $ & $\ 2S_{1/2},0,0\ $\\ \hline\hline
\mycell{18mm}{$\ 2P_{3/2},2,-1\ $} & 0 & 0 & 0 & $-\frac3{\sqrt2}\sep$ & 0 & 0 \\ 
\mycell{18mm}{$\ 2P_{3/2},1,-1\ $} & 0 & 0 & 0 & $\sqrt{\frac32}\sep$ & 0 & $-\sqrt6\sep$ \\ 
\mycell{18mm}{$\ 2P_{1/2},1,-1\ $} & 0 & 0 & 0 & $\sqrt3\sep$ & 0 & $\sqrt3\sep$ \\ 
\mycell{18mm}{$\ 2S_{1/2},1,-1\ $} & $\sqrt{\frac32}\sep$ & $\sqrt{\frac32}\sep$ & $\sqrt3\sep$ & 0 & $\sqrt3\sep$ & 0 \\ \hline
\end{tabular}\\[15pt]
\centerline{\bf (Table \ref{t:H2.D}f)}
\end{center}
\end{table}
\begin{table}
\begin{center}
\begin{tabular}{l||cccc|c|}
& \mycell{18mm}{$\ 2P_{3/2},2,-1\ $} & $\ 2P_{3/2},1,-1\ $ & $\ 2P_{1/2},1,-1\ $ & $\ 2S_{1/2},1,-1\ $ & $\ 2P_{3/2},2,-2\ $\\ \hline\hline
\mycell{18mm}{$\ 2P_{3/2},2,-1\ $} & 0 & 0 & 0 & 
$\frac{3}{{\sqrt{2}}}\sez$ & 0\\  
\mycell{18mm}{$\ 2P_{3/2},1,-1\ $} & 0 & 0 & 0 & 
${\sqrt{\frac{3}{2}}}\sez$ & 0\\  
\mycell{18mm}{$\ 2P_{1/2},1,-1\ $} & 0 & 0 & 0 & 
${\sqrt{3}}\sez$ & 0\\  
\mycell{18mm}{$\ 2S_{1/2},1,-1\ $} & 
$\frac{3}{{\sqrt{2}}}\sez$ & 
${\sqrt{\frac{3}{2}}}\sez$ & 
${\sqrt{3}}\sez$ & 0 & $3\sem$\\ \hline
\mycell{18mm}{$\ 2P_{3/2},2,-2\ $} & 0 & 0 & 0 & $-3\sep$ & 0\\ \hline
\end{tabular}\\[15pt]
\centerline{\bf (Table \ref{t:H2.D}g)}
\end{center}
\end{table}}

{
\begin{table}
\begin{center}
\caption{The suitably normalised magnetic dipole operator $\uvec{\mu}/\mu_B$ for the $n=2$ states of hydrogen, where $\mu_B = e\hbar/(2m_e)$ is the Bohr magneton and $g=2.002319304(76)$ is the Land\'{e} factor of the electron \cite{Moh05}.}\label{t:H2.Mu}
\vskip10pt
\begin{tabular}{l||c|cccc|}
& \mycell{18mm}{$\ 2P_{3/2},2,2\ $} & $\ 2P_{3/2},2,1\ $ & $\ 2P_{3/2},1,1\ $ & $\ 2P_{1/2},1,1\ $ & $\ 2S_{1/2},1,1\ $\\ 
%& $2,2$ & $2,1$ & $1,1$ & $1,1$ & $1,1$\\ 
\hline\hline
\mycell{18mm}{$\ 2P_{3/2},2,2\ $} & $-\frac{g+2}{2}\sez$ & $-\frac{\sqrt2(g+2)}{4}\sem$ & $\frac{\sqrt2(g+2)}{4\sqrt3}\sem$ & $-\frac{g-1}{\sqrt3}\sem$ & 0\\ \hline
\mycell{18mm}{$\ 2P_{3/2},2,1\ $} & $\frac{\sqrt2(g+2)}{4}\sep$ & $-\frac{g+2}{4}\sez$ & $-\frac{g+2}{4\sqrt3}\sez$ & $-\frac{g-1}{\sqrt6}\sez$ & 0\\
\mycell{18mm}{$\ 2P_{3/2},1,1\ $} & $-\frac{\sqrt2(g+2)}{4\sqrt3}\sep$ & $-\frac{g+2}{4\sqrt3}\sez$ & $-\frac{5(g+2)}{12}\sez$ & $\frac{g-1}{3\sqrt2}\sez$ & 0\\
\mycell{18mm}{$\ 2P_{1/2},1,1\ $} & $\frac{g-1}{\sqrt3}\sep$ & $-\frac{g-1}{\sqrt6}\sez$ & $\frac{g-1}{3\sqrt2}\sez$ & $\frac{g-4}6\sez$ & 0\\
\mycell{18mm}{$\ 2S_{1/2},1,1\ $} & 0 & 0 & 0 & 0 & $-\frac{g}2\sez$\\ \hline
\end{tabular}\\[15pt]
\centerline{\bf (Table \ref{t:H2.Mu}a)}\vskip15pt
% \end{table}
% \begin{table}
\begin{tabular}{l||cccccc|}
& \mycell{18mm}{$\ 2P_{3/2},2,0\ $} & $\ 2P_{3/2},1,0\ $ & $\ 2P_{1/2},1,0\ $ & $\ 2S_{1/2},1,0\ $ & $\ 2P_{1/2},0,0\ $ & $\ 2S_{1/2},0,0\ $\\ \hline\hline
\mycell{18mm}{$\ 2P_{3/2},2,1\ $} & $-\frac{\sqrt3(g+2)}4\sem$ & $\frac{g+2}{4\sqrt3}\sem$ & $-\frac{g-1}{\sqrt6}\sem$ & 0 & 0 & 0\\ 
\mycell{18mm}{$\ 2P_{3/2},1,1\ $} & $-\frac{g+2}{12}\sem$ & $-\frac{5(g+2)}{12}\sem$ & $-\frac{\sqrt2(g-1)}{6}\sem$ & 0 & $-\frac{\sqrt2(g-1)}{3}\sem$ & 0\\ 
\mycell{18mm}{$\ 2P_{1/2},1,1\ $} & $\frac{\sqrt2(g-1)}{6}\sem$ & $-\frac{\sqrt2(g-1)}{6}\sem$ & $\frac{g-4}{6}\sem$ & 0 & $-\frac{g-4}{6}\sem$ & 0\\ 
\mycell{18mm}{$\ 2S_{1/2},1,1\ $} & 0 & 0 & 0 & $-\frac{g}2\sem$ & 0 & $\frac{g}2\sem$\\ \hline
\end{tabular}\\[15pt]
\centerline{\bf (Table \ref{t:H2.Mu}b)}
\end{center}
\end{table}
\begin{table}
\begin{center}
\begin{tabular}{l||cccc|}
& \mycell{18mm}{$\ 2P_{3/2},2,1\ $} & $\ 2P_{3/2},1,1\ $ & $\ 2P_{1/2},1,1\ $ & $\ 2S_{1/2},1,1\ $\\ \hline\hline
\mycell{18mm}{$\ 2P_{3/2},2,0\ $} & $\frac{\sqrt3(g+2)}4\sep$ & $\frac{g+2}{12}\sep$ & $-\frac{\sqrt2(g-1)}{6}\sep$ & 0 \\ 
\mycell{18mm}{$\ 2P_{3/2},1,0\ $} & $-\frac{g+2}{4\sqrt3}\sep$ & $\frac{5(g+2)}{12}\sep$ & $\frac{\sqrt2(g-1)}{6}\sep$ & 0 \\ 
\mycell{18mm}{$\ 2P_{1/2},1,0\ $} & $\frac{g-1}{\sqrt6}\sep$ & $\frac{\sqrt2(g-1)}{6}\sep$ & $-\frac{g-4}{6}\sep$ & 0 \\ 
\mycell{18mm}{$\ 2S_{1/2},1,0\ $} & 0 & 0 & 0 & $\frac{g}2\sep$ \\ 
\mycell{18mm}{$\ 2P_{1/2},0,0\ $} & 0 & $\frac{\sqrt2(g-1)}{3}\sep$ & $\frac{g-4}6\sep$ & 0 \\ 
\mycell{18mm}{$\ 2S_{1/2},0,0\ $} & 0 & 0 & 0 & $-\frac{g}2\sep$ \\ \hline
\end{tabular}\\[15pt]
\centerline{\bf (Table \ref{t:H2.Mu}c)}\vskip15pt
% \end{table}
% \begin{table}
\begin{tabular}{l||cccccc|}
& \mycell{18mm}{$\ 2P_{3/2},2,0\ $} & $\ 2P_{3/2},1,0\ $ & $\ 2P_{1/2},1,0\ $ & $\ 2S_{1/2},1,0\ $ & $\ 2P_{1/2},0,0\ $ & $\ 2S_{1/2},0,0\ $\\ \hline\hline
\mycell{18mm}{$\ 2P_{3/2},2,0\ $} & 0 & 
$-\frac{g+2}{6}\sez$ & 
$-\frac{\sqrt2(g-1)}{3}\sez$ & 0 & 0 & 0\\
\mycell{18mm}{$\ 2P_{3/2},1,0\ $} & 
$-\frac{g+2}{6}\sez$ & 0 & 0 & 0 & 
$-\frac{{\sqrt{2}}(g-1)}{3}\sez$ & 0\\ 
\mycell{18mm}{$\ 2P_{1/2},1,0\ $} & 
$-\frac{{\sqrt{2}}(g-1)}{3}\sez$ & 0 & 0 & 0 & 
$\frac{g-4}{6}\sez$ & 0\\ 
\mycell{18mm}{$\ 2S_{1/2},1,0\ $} & 0 & 0 & 0 & 0 & 0 & 
$-\frac{g}{2}\sez$\\ 
\mycell{18mm}{$\ 2P_{1/2},0,0\ $} & 0 & 
$-\frac{{\sqrt{2}}(g-1)}{3}\sez$ & 
$\frac{g-4}{6}\sez$ & 0 & 0 & 0\\
\mycell{18mm}{$\ 2S_{1/2},0,0\ $} & 
0 & 0 & 0 & 
$-\frac{g}{2}\sez$ & 
0 & 0\\ \hline
\end{tabular}\\[15pt]
\centerline{\bf (Table \ref{t:H2.Mu}d)}
\end{center}
\end{table}
\begin{table}
\begin{center}
\begin{tabular}{l||cccc|}
& \mycell{18mm}{$\ 2P_{3/2},2,-1\ $} & $\ 2P_{3/2},1,-1\ $ & $\ 2P_{1/2},1,-1\ $ & $\ 2S_{1/2},1,-1\ $\\ \hline\hline
\mycell{18mm}{$\ 2P_{3/2},2,0\ $} & 
$-\frac{\sqrt3(g+2)}{4}\sem$ & 
$\frac{g+2}{12}\sem$ & 
$-\frac{\sqrt2(g-1)}{6}\sem$ & 0\\
\mycell{18mm}{$\ 2P_{3/2},1,0\ $} & 
$-\frac{g+2}{4 \sqrt{3}}\sem$ & 
$-\frac{5 (g+2)}{12}\sem$ & 
$-\frac{\sqrt2(g-1)}{6}\sem$ & 0\\ 
\mycell{18mm}{$\ 2P_{1/2},1,0\ $} & 
$\frac{g-1}{\sqrt{6}}\sem$ & 
$-\frac{\sqrt2(g-1)}{6}\sem$ & 
$\frac{g-4}{6}\sem$ & 0\\ 
\mycell{18mm}{$\ 2S_{1/2},1,0\ $} & 0 & 0 & 0 & 
$-\frac{g}{2}\sem$\\ 
\mycell{18mm}{$\ 2P_{1/2},0,0\ $} & 0 & 
$\frac{\sqrt2(g-1)}{3}\sem$ & 
$\frac{g-4}{6}\sem$ & 0\\ 
\mycell{18mm}{$\ 2S_{1/2},0,0\ $} & 0 & 0 & 0 & 
$-\frac{g}{2}\sem$\\ \hline
\end{tabular}\\[15pt]
\centerline{\bf (Table \ref{t:H2.Mu}e)}\vskip15pt
% \end{table}
% \begin{table}
\begin{tabular}{l||cccccc|}
& \mycell{18mm}{$\ 2P_{3/2},2,0\ $} & $\ 2P_{3/2},1,0\ $ & $\ 2P_{1/2},1,0\ $ & $\ 2S_{1/2},1,0\ $ & $\ 2P_{1/2},0,0\ $ & $\ 2S_{1/2},0,0\ $\\ \hline\hline
\mycell{18mm}{$\ 2P_{3/2},2,-1\ $} & 
$\frac{\sqrt3(g+2)}{4}\sep$ & 
$\frac{g+2}{4 \sqrt{3}}\sep$ & 
$-\frac{g-1}{\sqrt{6}}\sep$ & 0 & 0 & 0 \\ 
\mycell{18mm}{$\ 2P_{3/2},1,-1\ $} & 
$-\frac{g+2}{12}\sep$ & 
$\frac{5 (g+2)}{12}\sep$ & 
$\frac{\sqrt2(g-1)}{6}\sep$ & 0 & 
$-\frac{\sqrt2(g-1)}{3}\sep$ & 0 \\ 
\mycell{18mm}{$\ 2P_{1/2},1,-1\ $} & 
$\frac{\sqrt2(g-1)}{6}\sep$ & 
$\frac{\sqrt2(g-1)}{6}\sep$ & 
$-\frac{g-4}{6}\sep$ & 0 & 
$-\frac{g-4}{6}\sep$ & 0 \\ 
\mycell{18mm}{$\ 2S_{1/2},1,-1\ $} & 0 & 0 & 0 & 
$\frac{g}{2}\sep$ & 0 & $\frac{g}{2}\sep$ \\ \hline
\end{tabular}\\[15pt]
\centerline{\bf (Table \ref{t:H2.Mu}f)}
\end{center}
\end{table}
\begin{table}
\begin{center}
\begin{tabular}{l||cccc|c|}
& \mycell{18mm}{$\ 2P_{3/2},2,-1\ $} & $\ 2P_{3/2},1,-1\ $ & $\ 2P_{1/2},1,-1\ $ & $\ 2S_{1/2},1,-1\ $ & $\ 2P_{3/2},2,-2\ $\\ \hline\hline
\mycell{18mm}{$\ 2P_{3/2},2,-1\ $} & 
$\frac{g+2}{4}\sez$ & 
$-\frac{g+2}{4{\sqrt{3}}}\sez$ & 
$-\frac{g-1}{{\sqrt{6}}}\sez$ & 
0 & $-\frac{\sqrt2(g+2)}{4}\sem$\\
\mycell{18mm}{$\ 2P_{3/2},1,-1\ $} & 
$-\frac{g+2}{4{\sqrt{3}}}\sez$ & 
$\frac{5(g+2)}{12}\sez$ & 
$-\frac{g-1}{3{\sqrt{2}}}\sez$ & 
0 & $-\frac{\sqrt2(g+2)}{4\sqrt3}\sem$\\ 
\mycell{18mm}{$\ 2P_{1/2},1,-1\ $} & 
$-\frac{g-1}{{\sqrt{6}}}\sez$ & 
$-\frac{g-1}{3{\sqrt{2}}}\sez$ & 
$-\frac{g-4}{6}\sez$ & 
0 & $\frac{g-1}{\sqrt3}\sem$\\ 
\mycell{18mm}{$\ 2S_{1/2},1,-1\ $} & 
0 & 0 & 0 & 
$\frac{g}{2}\sez$ & 0\\ \hline
\mycell{18mm}{$\ 2P_{3/2},2,-2\ $} & $\frac{\sqrt2(g+2)}{4}\sep$ & $\frac{\sqrt2(g+2)}{4\sqrt3}\sep$ & $-\frac{g-1}{\sqrt3}\sep$ & 0 & 
$\frac{2 + g}{2}\sez$\\ \hline
\end{tabular}\\[15pt]
\centerline{\bf (Table \ref{t:H2.Mu}g)}
\end{center}
\end{table}}
\vfill
%\end{turnpage}

\twocolumn

%%====================================================================================================

\bibliographystyle{epj}
\bibliography{myapvbib}
%\bibliography{epj}

\end{document}